\documentclass[preprint2]{aastex}

\shorttitle{CG~J1720-67.8: Optical and IR Properties}

\begin{document}

\title{CG~J1720-67.8: a detailed analysis of optical and infrared properties of a new
ultracompact group of galaxies\footnote{Partly based on data collected at the European 
Southern Observatory, La Silla, Chile, Proposal 63.N-0737.}}

\author{S. Temporin and R. Weinberger}
\affil{Institut f\"ur Astrophysik, Leopold-Franzens-Universit\"at Innsbruck, 
Technikerstra\ss e 25, A-6020 Innsbruck, Austria}
\email{giovanna.temporin@uibk.ac.at, ronald.weinberger@uibk.ac.at}

\author{G. Galaz}
\affil{Departamento de Astronom\'{\i}a y Astrof\'{\i}sica, Pontificia Universidad Cat\'olica de
Chile,\\
Casilla 306, Santiago 22, Chile}
\email{ggalaz@astro.puc.cl}

\and

\author{F. Kerber}
\affil{Space Telescope European Coordinating Facility, European Southern 
Observatory, Karl-Schwarzschild-Stra\ss e 2, D-85748 Garching, Germany}
\email{fkerber@eso.org}

\begin{abstract}
We present here optical spectroscopy and BVRJHK$\rm _s$ photometry of the 
recently discovered ultra-compact group of galaxies CG~J1720-67.8.

This work represents a considerable extension of the preliminary
results we presented in a previous paper. 
Despite the complicated morphology of the group, a quantitative morphological classification
of the three brightest members of the group is attempted based on photometric analysis.
We find that one galaxy
is consistent with a morphological type S0, while the other two 
are most probably late-type spirals that are already losing their identity due to 
the interaction process. 
Information on the star formation activity and dust content derived from both spectroscopic
data and optical and near-infrared colors are complemented with a reconstruction of 
far-infrared (FIR) maps from IRAS raw data.
Enhanced star formation activity is revealed in all the group's members, including
the early-type galaxy and the extended tidal tail,
along which several tidal dwarf galaxy candidates are identified.
The metallicity of the gaseous component is investigated and photoionization
models are applied to the three main galaxies of the group, while a detailed study
of the tidal dwarf candidates will appear in a companion paper. Subsolar metal 
abundances are found for all the three galaxies, the highest values being shown by 
the early-type galaxy (Z $\sim$ 0.5 Z$_{\sun}$).
\end{abstract}


\section{INTRODUCTION}

The fate of compact groups of galaxies (CGs) seems to be established:
their small galaxy-galaxy separation and low velocity dispersion facilitate
the merging of their member galaxies leading to the final collapse into a 
bright field elliptical.
This picture, in agreement with the hierarchical galaxy formation theory
\citep{la90,rhs97,ka99} is supported by N-body 
simulations (e.g. Barnes 1990) and by recent observations both in the 
optical and in the X-ray domain.
Several isolated elliptical galaxies surrounded by an X-ray halo and by a 
population of dwarf galaxies have been suggested as relics of CGs 
\citep{po94,mz99,vi99,jpf00}. 
HST observations of ultraluminous IR galaxies
revealed that some of them are the product of CGs merging \citep{bo00}.

Nevertheless, in spite of the numerous studies of CGs and the ongoing surveys 
aiming at the identification of CGs' relics, the transition phase from a CG to 
an elliptical has not yet been fully analyzed and understood, due to the 
difficulty of finding such evolved systems.
This sparse coverage of transitional objects is why we think it worthwhile 
to concentrate our attention on a recently discovered, very compact 
galaxy group, whose properties, as described in \citet[hereafter Paper I]{wtk99}, 
are indicative of a very advanced evolutionary phase.
We briefly summarize in the following the main results achieved in Paper I:

CG~J1720-67.8 (z=0.045) is an isolated group of three galaxies with an 
associated arc-shaped luminous tidal feature at whose ends two brightness enhancements 
are present. These are likely forming (or already formed?) tidal dwarf galaxies (TDGs).
Several other knots are detected along the tidal ``arc'' and at the tip
of a small tidal tail departing from the second brightest galaxy.
The group is one of the most compact known so far. The median 
projected separation of the four brightest objects, 6.9 kpc\footnote{Throughout 
this paper we assume H$_0$ = 75 km s$^{-1}$ Mpc$^{-1}$}, and the low velocity
dispersion, $\sigma$ = 67 km s$^{-1}$, facilitate the 
interaction processes inside the group as evidenced by its peculiar morphology.
Optical emission lines were detected in the
four brightest objects, in a portion of the arc-like tidal feature, and 
in a small knot located at the end of a faint tail. They showed 
emission line ratios typical of \ion{H}{2} or starburst-nucleus-galaxies (SBNG). 
The H$\alpha$ luminosity, 
after correction for the high internal extinction derived from the Balmer 
decrement, actually revealed a particularly enhanced star formation activity 
possibly triggered by galaxy interactions.

In this work we make use of new photometric and spectroscopic data to perform a more
detailed study of the three brightest galaxies of CG~J1720-67.8 and we extend our analysis
to the southern part of the group; details about the remaining group's members are
given in a companion paper (Temporin et al. 2002).
Our new data show that the star formation activity is present all over  
the group and display the presence of diffuse light around the CG, as well as a 
number of candidate dwarf galaxies in its surroundings.
A quantitative morphological classification and a broad-band color inspection indicate that
one of the two brightest galaxies (no. 2, see Fig.~\ref{Rgroup}), despite its moderate 
star formation rate  
($\sim$ 0.7 M$_{\odot}$ yr$^{-1}$, Paper~I), is an early-type galaxy (most probably an S0),
reddened by a considerable amount of dust. 

The analysis of raw IRAS 
data by means of high resolution IRAS maps obtained with the 
Maximum Entropy Method appears to offer an answer to the puzzling question posed 
in Paper~I regarding the lack of IRAS cataloged sources despite beeing in such a
dusty star-forming object.
Finally, thanks to the full coverage of the group members with radial velocity 
measurements, we can now estimate the mass of the group, under the assumption 
that it is a virialized system. 

The present paper is organized as follows: observations and data processing are 
described in Section 2; color indices and absolute photometry as well as 
a description of the adopted galaxy fitting techniques and a discussion of the photometric 
results are presented in Section 3. The spectral analysis of the group, 
together with photoionization models for the three brightest galaxies, is presented in Section 4,
where considerations on the dust content of the group based on IRAS data are expressed as well,
and an estimate of the group's virial mass is provided. Finally
the results are summarized and discussed in Section 5.

\section{THE DATA}

The present work is based on new optical and near infrared (NIR) data 
and on the data already presented in Paper I, as well.
A summary of all the observations is given in Table~\ref{obslog}.
In the following we describe separately the data acquisition and reduction
of optical and NIR data.

\placetable{obslog}

\subsection{Optical Data}

New BVR imaging and Multi Object Spectroscopy (MOS) frames were acquired 
in June 1999 at the ESO 3.6 m telescope
in La Silla equipped with EFOSC2 and the Loral Lesser CCD\#40, whose
15 $\mu$-sized pixels gave a spatial scale of 0$^{\prime\prime}$.157 pixel$^{-1}$ on
a 5$^{\prime}$.4 $\times$ 5$^{\prime}$.4 field of view. 
The seeing varied between 1$^{\prime\prime}$.3 and 1$^{\prime\prime}$.8
during the observations.
The optical images suffered from unfavorable and unstable weather conditions and 
a \citet{la92} standard field could be observed only once in 
the three photometric bands, which was not sufficient for determining the calibration
coefficients. Therefore the photometric calibration had to
rely on mean calibration coefficients available at the La Silla Observatory
web-site\footnote{http://www.ls.eso.org/}. In particular we used zero points
and color terms given for the two photometric nights preceding and following 
our observations. Mean extinction coefficients were obtained from the measurements
performed at the La Silla site by the Geneva Observatory Photometric Group until 
1997\footnote{http://obswww.unige.ch/photom/extsil/}.
Therefore the photometric calibration must be taken with caution: although we 
performed some tests on its reliability (see Appendix A), we warn about possible 
systematic errors which could have been introduced.
Additionally the B-band images turned out to be heavily affected by intervening
clouds during the exposures, so for magnitude estimate purposes we preferred 
to use an acquisition image obtained by the ESO staff a few days before our 
observing run, under better (but still not ideal) weather conditions.

The use of the MOS technique allowed simultaneous acquisition of
spectra at
several positions on CG~J1720-67.8 using short slitlets 1$^{\prime\prime}$.5-wide, 
punched on 5 different masks. Typical exposures were 1800 s with every mask.
The actual wavelength interval varied depending on the position of the 
slitlets on the frame, in the range 3350 -- 7560 \AA\ given by grism \#11.
The spectral resolution was $\sim$ 19 \AA.

The usual reduction steps (bias subtraction, flat-fielding, cosmic-rays 
removal, and for spectra, additional wavelength calibration) were performed by 
means of standard IRAF\footnote{IRAF is distributed by the National Optical 
Astronomy Observatories, which are operated by the Association of Universities
for Research in Astronomy, Inc., under cooperative agreement with the 
National Science Foundation} packages. An additional defringing procedure was 
applied to R-band images due to the presence of a fringe-pattern 
produced by night-sky emission lines.
For this purpose, after verification of the constancy of the fringe pattern across the 
two nights of observation, the median of all the R-band, background
subtracted images, was used as a fringe-frame to be conveniently scaled and subtracted from
the R images of the galaxy group.
After alignment, the three R-band images were averaged in order to improve the signal-to-noise ratio.
In order to appreciate color gradients across
the galaxy group, B$-$V, V$-$R and B$-$R color maps were obtained after having matched the 
point-spread-functions (PSFs) of the BVR images through convolution 
with appropriate Gaussian functions.
The result will be analyzed in Section 3, where the
photometric analysis, carried out by use of the 
package GIM2D \citep{sim98,sim01} is discussed as well.

\placefigure{Rgroup}

The positions in the galaxy group, where spectra were obtained either in June 1999
or during previous observing runs, are labeled onto the deep R image in Fig.~\ref{Rgroup}.
MOS spectra were flux calibrated using a 300 s exposure of the standard star 
G-138-31 obtained during the same night through a 1$^{\prime\prime}$.5-wide long-slit.
One-dimensional spectra of the individual objects were extracted and corrected
for foreground galactic extinction. This correction was applied by assuming 
A$\rm _V$ = 0.290, a value given in NED\footnote{NASA/IPAC Extragalactic Database,
operated by the Jet Propulsion Laboratory, California Institute of Technology,
under contract with the National Aeronautics and Space Administration} as 
deduced from \citet{sch98} and in agreement with our Galactic extinction estimate 
given in Paper~I.
Spectral emission features were measured by 
Gaussian fitting with the IRAF task SPLOT. Deblending of H$\alpha$ + [\ion{N}{2}] 
triplets was achieved by using the MIDAS\footnote{The ESO-MIDAS system is
developed and maintained by the European Southern Observatory} task ALICE.
Where absorption lines could be identified, a template of the underlying 
stellar component was appropriately diluted to match the absorption features
 and subtracted from the original spectrum, following the
prescriptions by \citet{hfs93}, in order to minimize the contamination by absorption
lines on the emission-line measurements.
Emission line fluxes and radial velocities given in Tables~\ref{galflux} and 
\ref{tdgflux} are both from the present work and from Paper I\footnote{The measurements of
the spectrum of galaxy 2 and object no. 3 given here slightly differ from the ones given in Paper~I, 
because of a redefinition of the section of the bidimensional spectrum occupied by the objects.}. 

\placetable{galflux}
\placetable{tdgflux}

\subsection{NIR Data}

NIR images in the photometric bands J, H, and K-short (K$\rm _s$) were acquired in June 1999
under photometric conditions at the 1m Swope telescope
in Las Campanas, using a 256 $\times$ 256 pixel NICMOS3 HgCdTe array having 
a pixel size of 40 $\mu$m and giving a spatial scale of 0$^{\prime\prime}$.599 
pixel$^{-1}$ on a 2$^{\prime}$.5 $\times$ 2$^{\prime}$.5 field of view.
The H, J, and K$\rm _s$ filters are centered at 1.24, 1.65, and 2.16 $\mu$m 
respectively, and have bandwidths of 0.22, 0.30, and 0.33 $\mu$m.
The properties of this photometric system are described in detail in \citet{petal98}.
For the determination of the photometric calibration coefficients, six
standard stars from the catalog of \citet{petal98} were observed
(identification numbers: 9119, 9136, 9144, 9146, 9170, 9178).

The observational procedure as well as the reduction steps were analogous to
the ones described in \citet{ga00}.
Every exposure of both the target and the standard stars was split into a 
number of dithered images (see Table 1) depending on the magnitude of the 
objects, the sky brightness and the linearity regime of the detector.
Total exposure times of 1800 s in J, 2000 s in H, and 2500 s in K$\rm _s$ were 
achieved for CG~J1720-67.8, while the final integration time for the standard
stars was 50 s in each filter. The processing of the raw images with a 
reduction pipeline based on the DIMSUM\footnote{Deep Infrared Mosaicing
Software package, developed by P. Eisenhardt, M. Dickinson, A. Stanford, \& 
J. Ward, and available from 
ftp://iraf.noao.edu/iraf/contrib/dimsumV2/dimsum.tar.Z.}
package included dark correction, flat-fielding, coaddition, alignment, cosmic
rays rejection, registration, and sky subtraction.
The PSFs of the images were matched through convolution
with a circular Gaussian function and finally the infrared color maps
J$-$H, H$-$K$\rm _s$, and J$-$K$\rm _s$ were made.

\section{PHOTOMETRIC MEASUREMENTS}

\subsection{Color Indices}

The extreme compactness of the group suggests that it is in an advanced stage of 
evolution and we might expect to infer some insight into its evolutionary stage 
and into interaction-induced events, like enhancement of star formation and 
subsequent presence of young stellar populations, from a qualitative 
comparison of its appearance in different wavelength bands and from optical and 
NIR color maps. In Fig.~\ref{allbands} the sequence of 
BVRIHJK$\rm _s$ images is shown for comparison purposes, while in Fig.~\ref{allcolors} the 
optical and NIR color maps B$-$V, V$-$R, B$-$R, J$-$H, H$-$K$\rm _s$ and J$-$K$\rm _s$ 
show the relative colors of the different parts of the group, giving evidence of color gradients
across it. 
A diffuse envelope, a kind of halo, 
is present in all the optical images and in the I band, but it 
seems to be more prominent in V and R. This halo is possibly a mixture of stars, dust 
and excited gas stripped away from the parent galaxies and now trapped into the 
group's potential well: the presence of a common halo of optical light is 
one of the evidences of the physical binding among the members of a group.
\placefigure{allbands}
\placefigure{allcolors}

In the B$-$V and B$-$R color maps a color gradient is visible across
galaxy 2, whose nucleus is also significantly redder than all the other objects of the group. 
A less marked color gradient can be observed in galaxy 4. The color difference 
between the central parts of the two galaxies is much less evident in 
V$-$R. The bluest regions in the maps are occupied by galaxy 1, and objects 3, 9, 7, and 8. 
The whole tidal arc exhibits rather blue colors, too, and the emission knots 
along it are visible as even bluer regions. Finally, the ring-like structure to 
which object 11 belongs also shows blue colors, although there the signal-to-noise 
ratio (SNR) is very low. In the V$-$R color the diffuse halo appears to be reddish, with 
some bluer condensations, but again the SNR is too low to allow us to draw 
significant conclusions. The red blob besides the nucleus of galaxy 1, well
visible in all the color maps, is a foreground star projected onto the galaxy, as 
shown by its spectrum (see Section 4).

We interpret the characteristics described above as an 
indication of recent and/or ongoing star formation spread over most of the group's 
extent. Young stellar populations must be present in the bluest parts of the 
group, while a mixture of old and young populations might be present in galaxy 
2, though metallicity effects and dust extinction might play a role in the
observed color gradients (age-metallicity-dust degeneracy; see e.g. \citet{wo94}).

In the NIR images the tidal arc appears to gradually fade with increasing wavelengths, 
supporting the hypothesis that old stars stemming from the parent galaxies are 
present in it --along with a young population responsible for its blue color-- and 
contribute to the NIR light. The common halo becomes extremely faint 
in these passbands and is barely visible. Objects 3, 7, and 8 almost 
disappear in the K$\rm _s$ frame. 

Since NIR bands are considerably less affected by reddening, a comparison
between the optical color maps and the NIR ones (Fig.~\ref{allcolors})
is already helpful for disentangling dust extinction effects and to understand
where the dust is located in the group. In J$-$H and J$-$K$\rm _s$ maps,
the central part of galaxy 2 appears bluer than its surroundings, which is the
opposite of what is observed in optical colors. The same holds for galaxy 4, 
although to a smaller degree. These color gradients suggest that star formation is more concentrated 
in the nuclear regions of the two galaxies and their redder optical colors 
are due (at least partly) to extinction effects. Specifically, a considerable
amount of dust seems to be present in (or along our line of sight to) galaxy 2, 
while the dust
distribution in galaxy 4 seems to be less extended and more patchy. Note that 
all the regions appearing bluer in the J$-$H and J$-$K$\rm _s$ maps are redder in
H$-$K$\rm _s$. This trend is not surprising, because this color
has an opposite behavior with respect to the others and becomes redder 
during starburst phases (see Temporin \& Fritze-von Alvensleben 2002).
Such observations are in agreement with the spectroscopic results:
the high values of the Balmer decrement (Tables~\ref{galflux} and \ref{tdgflux}) are consistent 
with a large amount of internal extinction, although these values are necessarily referred only 
to the portions of galaxies entering the slit. 

\subsection{Optical and NIR Photometry}

Optical instrumental magnitudes were calibrated in 
the La Silla photometric system by the following equations:
\begin{equation}
 B = b + 25.954 - 0.278 X{\rm _B} + 0.030 (B-V)
\label{Btransf}
\end{equation}
\begin{equation}
 V = v + 25.974 - 0.140 X{\rm _V} + 0.0215 (B-V)
\label{Vtransf}
\end{equation}
\begin{equation}
R = r + 26.070 - 0.095 X{\rm _R} + 0.003 (V-R)
\label{Rtransf}
\end{equation}
where $B$, $V$, $R$ are calibrated magnitudes, $b$, $v$, $r$ are
sky-subtracted instrumental magnitudes normalized to 1 s exposure and 
$X{\rm _B}$, $X{\rm _V}$, $X{\rm _R}$ are the airmasses at the time of the observations.
 The surface brightness was found to be sky-background limited at the values 
25.64 mag arcsec$^{-2}$ in B, 25.28 mag arcsec$^{-2}$ in V, and 24.89 
mag arcsec$^{-2}$ in R by computing the surface brightness of a pixel having a 3$\sigma$
detection. The limiting magnitudes in the three bands were found to be B$_{\rm lim}$ = 25.05 mag, 
V$_{\rm lim}$ = 24.32 mag, and R$_{\rm lim}$ = 24.23 mag by referring to a 3$\sigma$ detection 
and taking into account the area of the seeing disk.

The photometric calibration in NIR bands was performed using the standard stars, which
were observed during the same night of the target at air masses ranging from 
1.32 to 1.07.
The calibration coefficients were determined through a uniformly weighted fit
of the following transformation equations:
\begin{equation}
J = J_0 + j -0.10 X{\rm _J}
\label{Jtransf}
\end{equation}
\begin{equation}
H = H_0 + h -0.04 X{\rm _H}
\label{Htransf}
\end{equation}
\begin{equation}
K = K_0 + k -0.08 X{\rm _K}
\label{Ktransf}
\end{equation}
where $J_0$, $H_0$, $K_0$ are the zero points of the photometric 
system, $X{\rm _J}$, $X{\rm _H}$, $X{\rm _K}$ are the air masses, and $j$, $h$, $k$ are the 
instrumental magnitudes of the standard stars. The extinction coefficients 
are the ones given in \citet{petal98}. Color terms were not included 
because they were smaller than the magnitude errors.
The $rms$ from the fits to the transformation equations was $\sim$ 0.02 -- 0.03,
which gives an estimate of the internal error of the photometric calibration.
The obtained zero points are: $J_0$ = 19.79 $\pm$ 0.01, $H_0$ = 19.34 $\pm$ 
0.01, and $K_0$ = 18.86 $\pm$ 0.01. The limiting surface brightnesses and 
magnitudes are: 
$\mu_{\rm Jlim}$ = 
22.21 mag arcsec$^{-2}$, $\mu_{\rm Hlim}$ = 20.92 mag arcsec$^{-2}$, $\mu_{\rm 
Klim}$ = 20.18 mag arcsec$^{-2}$, J$_{\rm lim}$ = 21.71 mag, H$_{\rm lim}$ = 
20.24 mag, and K$_{\rm lim}$ = 19.65 mag.

\subsection{Galaxy Fitting Procedure}

Photometry of the three galaxies was achieved through a bi-dimensional fitting of the galaxy images.
For this purpose we used GIM2D
\citep{sim98,sim01}, an IRAF package written to perform an automated bulge-disk
surface brightness profile decomposition of low signal-to-noise images of
distant galaxies. The program gives as output, besides the value of the best-fit parameters
and asymmetry indices, an image of the galaxy model and a residual image obtained by subtracting
the model from the original image. 
Although GIM2D does not allow
to follow changes in ellipticity, center position, and position angle of the isophotes
across a galaxy, it has the considerable advantage of including a PSF deconvolution 
in the bi-dimensional fit of the galaxy image.
The small-sized galaxies of CG~J1720-67.8 --having semi-major axis of the outermost visible isophotes
in our images ranging from 4 to 8 arcsec-- 
could have their profiles significantly affected by the PSF. Therefore a bulge-disk
decomposition that takes into account this effect is the most suitable to a quantitative
morphological classification. 

This image fitting method has already been successfully applied
by \citet{tr01} to a sample of galaxies belonging to poor galaxy groups 
and spanning a wide range in morphological type. They found that the ratio
between the bulge and the total luminosity (B/T) can be used on average as a robust morphology 
indicator to discriminate between early-type bulge-dominated (B/T $>$ 0.4) and late-type
disk-dominated (B/T $<$ 0.4) galaxies, although for individual galaxies B/T can change in 
either direction by as much as 59\%.

For the application of GIM2D to CG~J1720-67.8 we followed the prescriptions given
in \citet{sim01}, as well as in the related web
site\footnote{http://nenuphar.as.arizona.edu/simard/gim2d/gim2d.html}.
PSF images in the six photometric bands to be used during the galaxy fitting procedure
were obtained with the DAOPHOT \citep{sh88} package inside IRAF.
The isophotal area of the galaxies was defined 
by means of the galaxy photometry package SExtractor V2.2.2 \citep{ba96}
with a minimum deblending contrast parameter of 0.0001 in order to separate
the three galaxies from one each other and from the tidal arc. 
The fit uses an exponential law for the disk, while a classical de Vaucouleurs $r^{1/4}$-law 
\citep{dv48} or
a S\'ersic profile \citep{se68} can be chosen for the bulge. In the last case, the
$n$ index of the S\'ersic law is one of the fitting parameters.
We attempted fits of the three galaxies both with a de Vaucouleurs bulge plus 
exponential disk, and with a S\'ersic law plus exponential disk.
Comparably good fits --as it was judged from the reduced $\chi^2$ of the fits and the total residual
light R$_T$-- were obtained for all the three galaxies with both kinds of law for the bulge with no
appreciable change in the total integrated luminosity of the galaxies.

Nevertheless, the de Vaucouleurs profile causes an oversubtraction of the region around the center of
galaxy 1 in the residual image and a failure in the detection of the disk of this galaxy in the B band
(B/T = 1), while a better residual image and a B/T = 0.4 are obtained with the S\'ersic
profile. The best fit with the latter gives a  S\'{e}rsic index $n$ = 1.3, thus  
approaching an exponential law also for the bulge.
Similarly the best fit of galaxy 4 with a S\'ersic bulge plus exponential disk led to a S\'ersic index
$n$ = 0.8. Actually in recent years both observational studies and numerical simulations have
provided growing indications that bulges of (especially late-type) spiral galaxies follow an
exponential law rather than the classical de Vaucouleurs one \citep{as94,a98,ca01}.
The S\'ersic index for galaxy 2, $n$ = 4.0, indicates that the de Vaucouleurs law is a good
representation of its bulge profile.

The complex morphology of CG~J1720-67.8 requires particular attention in disentangling the light
coming from different objects when measuring their magnitudes. Owing to the compactness of the group,
the three brightest galaxies are partly overlapping (at least in projection on the plane of the sky).
In order to deal with this problem, we have adopted the iterative fitting procedure described below.

In every image we started by fitting galaxies 1 and 4 and subtracting their models from the original
frame. Galaxy 2 was fitted on the residual image and its model was subtracted from the original one.
A new fit of galaxies 1 and 4 was performed on the new residual frame. After the subtraction of their
models, a second fit of galaxy 2 was obtained. Further iterations of this fitting/subtraction procedure
did not show significant changes in the result.
Fig.~\ref{bmodel} makes clear the importance of the above explained procedure: in the left panel (original
frame $-$ galaxy 2) the outskirts of galaxy 4 appear to overlap galaxy 2 up to its center; similarly in
the right panel (original frame $-$ galaxies 1 and 4) galaxy 2 extends to the center
of galaxy 4 and significantly overlaps with galaxy 1.

\placefigure{bmodel}

We show in Fig.~\ref{gim2dres} the residual image of the group after subtraction of the best fit 
models of all the three galaxies.
\placefigure{gim2dres}
A knot of residual light remained just below the center of galaxy 2, as well as 
the bridge connecting galaxy 1 to galaxy 2, already evident in the original image,
and another knot near the center of galaxy 1.
Galaxy 4 shows internal features and blobs reminiscent of a knotty spiral structure and going
from the upper part of the tidal arc through the galaxy center to object no. 5.
Supported also by the enhanced star formation activity of the galaxies (see Section 4),
we interpret these residual structures as \ion{H}{2} regions. Thus galaxies 2 and 4 appear 
to have \ion{H}{2} regions located next to the nucleus, in agreement with the spectral 
classification in Section 4.3.

\subsection{Total Magnitudes and Structural Parameters} 

With the fit of the surface brightness distribution of the galaxies we obtained a measure
of the total integrated galaxy light, the B/T ratio and other structural parameters,
namely the effective radius (i.e. the half-light
radius of the bulge component) $r_e$, the disk scale-length $h_D$, the bulge ellipticity $e$,
the position angles of bulge and disk components $\phi_b$ and $\phi_d$, 
the disk inclination $i$ and the half-light radius $r_{1/2}$ (i.e. the radius inside which half
of the light of the whole galaxy is contained). 
Additionally the R$_T$ parameter giving the total residual light after the galaxy model subtraction
allows an estimate of the goodness of the fit, while the asymmetry index R$_A$, measured at
2$r_{1/2}$,  gives an indication of the presence of asymmetric structures in the galaxies
\citep{tr01}.

The total magnitudes and colors obtained from the total integrated galaxy light given by GIM2D, after
calibration according to equations \ref{Btransf} to \ref{Ktransf}, are listed in Table~\ref{mag}. 
The structural
parameters of the galaxies --given by the best-fit with a de Vaucouleurs bulge plus an exponential 
disk-- are listed in Table~\ref{gim2dparam} along with the reduced-$\chi^2$ of the fits.

\placetable{mag}
\placetable{gim2dparam}

Column (1) of Table~\ref{mag} lists the observed magnitudes corrected for Galactic extinction 
by means of the extinction values given in NED, based on Schlegel et al. (1998).
An inclination correction taking into account the internal extinction has been calculated 
as a function of the morphological type T and of the B-band disk inclination $i$\footnote{The B-band 
disk inclination was used instead of the ratio R$_{25}$ between the major and minor diameters at the
$\mu_{\rm B}$ = 25 mag arcsec$^{-2}$ isophote, which could not be measured.}
(Table~\ref{gim2dparam}) following the
instructions given in the introduction to the Third Reference
Catalogue of Bright Galaxies \citep[RC3]{dv91}.
Based on our quantitative morphological analysis (previous section) and on a comparison of the
galaxy colors with the optical colors as a function of the morphological type T given by \citet{bw95},
we used T = 6 (i.e. Scd) for galaxy 1, T = 0 (i.e. S0) for galaxy 2, and T = 5 (i.e. Sc) for galaxy 4.
The internal extinction in the other bands was calculated from the 
B-band one, A$_{\rm B}$, and the color excess E(B$-$V) by applying the \citet{ca89} 
extinction law and calculating the selective extinction R$_{\rm V}$ according to equation (56) of the
RC3. These extinction values and color excesses are listed in column (2) of Table~\ref{mag} for each
galaxy, while face-on magnitudes and colors are given in column (3).
However the inclination correction is based on the assumption
that the galaxies contain the standard amount of dust typical of
galaxies of similar morphological type. Since there are indications of the presence
of a considerable amount of extinction in CG~J1720-67.8, in some cases the correction 
calculated above might be a lower limit. 

The results on the bulge-disk decomposition 
(see previous subsection and Table~\ref{gim2dparam}) 
could support the hypothesis that galaxy 1 is a late-type spiral, which has lost its outer
layers during the interaction with its companions and has remained with an exponential
bulge that dominates its light.
Nonetheless, as noticed by \citet{tr01}, a B/T $>$ 0.4 for galaxies
with strong [\ion{O}{2}] emission -- and therefore with enhanced star
formation, as is the case for galaxy 1 (see Section 4)-- could indicate 
the presence of central star forming regions rather than a true bulge.
Galaxy 4 can be classified as a late-type spiral (Sc), as well.
We note that, except for the internal structures revealed in the residual images, no traces of
spiral arms remain visible in this galaxy. We argue that its spiral structure might have
been destroyed by the interaction and could have been strewn into the adjacent arc-like 
tidal feature.

The bulge of galaxy 2 dominates its light, as already expected from a visual inspection of
the images, and has a de Vaucouleurs profile. However a disk component is present as well.
We finally classify galaxy 2 as an S0.

The sources of error in the total magnitudes 
are multiple and difficult to evaluate, in particular due to the
iterative procedure applied to disentangle the fluxes from partially
overlapping galaxies. Since GIM2D calculates a lower and an upper limit of the total galaxy 
flux, together with its best-fit value, we used the semi-difference of these limits to
have an indication of the magnitude error related to the fitting procedure.
We found that this error is ranging from $\sim$ 0.02 to $\sim$ 0.1 mag in the NIR, 
while it is $\sim$ 0.01 mag or lower in the optical bands.
Therefore in the case of optical magnitudes we consider the calibration error 
($\lesssim$ 0.05 mag, see Appendix) as the dominating one. 

The asymmetry index R$_A$ (Table~\ref{gim2dparam}) are considerably high for all the
three galaxies, except in the H and K$_{\rm s}$ bands.
\citet{tr01} adopt R$_A$ $\geq$ 0.05 as the lower limit for classifying a galaxy as
having high asymmetry.
A high value of R$_A$ can result from asymmetrically distributed \ion{H}{2} regions
and/or interaction-induced features (e.g. tidal distortions). 
After an inspection of the residual images, we interpret the R$_A$ values of the galaxies 
of CG~J1720-67.8 as an effect of tidal interactions, which are probably responsible for both
morphological distortions and the presence of asymmetrically distributed regions
of star formation.  

\section{RESULTS AND DISCUSSION}
 
\subsection{Widespread Emission Line Activity}

CG~J1720-67.8 is characterized, as already mentioned, by line emission 
over its spatial extent. Several parts of the group were spectroscopically 
sampled, as shown from the labeling in Fig.~\ref{Rgroup}. 
Besides the spectra analyzed in Paper~I, new spectra of galaxies 1 and 4 were acquired at a
different position angle (P.A. = 90\degr).
The wider wavelength range of the new spectrum allowed the detection of the 
[\ion{O}{2}] $\lambda$ 3727 line in galaxy 4, whose general spectral properties are in 
good agreement with those found in Paper I. For galaxy 1 a new spectrum across the nucleus
(1a) and one across the knot 1b were obtained. While the latter was found to be
a late-type foreground
star projected onto the galaxy, weak emission lines from the underlying galaxy 
were detected and allowed a measure of the radial velocity. Therefore three values of 
radial velocity, related to three different portions of the galaxy were available and 
we adopted the weighted mean as the radial velocity of galaxy 1.
The long-slit spectrum of galaxy 1 exhibits a secondary peak, as it is visible from the
H$\alpha$ profile along the slit (Fig.~\ref{Haprof}), which was used to identify and
extract the individual objects. Therefore we extracted two spectra of galaxy 1 at 
P.A. = 130\degr, namely one including both the peaks, labeled 1$\ast$, and one 
including only the main peak, labeled 1, and corresponding to the north-western side
of the galaxy. The one-dimensional spectra of the three galaxies, both from Paper~I and
from the present work, are shown in Figs.~\ref{galspec1} and \ref{galspec2}.
\placefigure{Haprof}
\placefigure{galspec1}
\placefigure{galspec2}

The additional objects for which spectra were obtained have redshifts in agreement
with the previously studied group's members. They are mainly located along the 
tidal arc (see Fig.~\ref{Rgroup}) and are possible tidal dwarf candidates 
(for further details see Temporin et al. 2002). 
The heliocentric radial velocities of the group's members listed
in Table~\ref{radvel} have been obtained from position measurements of their emission lines.
The value given for object no. 11 rests on the only detected emission line, whose wavelength
corresponds to the redshifted line [\ion{O}{2}] $\lambda$ 3727, at a redshift in agreement 
with that of the galaxy group. Radial velocities of objects 2, 3, 4, 5, and 6 are from Paper~I
and their internal errors were derived following the method described by \citet{co99}, which takes
into account the signal-to-noise ratio (SNR) of the emission lines.
This method could not be applied to the short-slit MOS spectra. As an alternative the velocity error
$\Delta v$ associated with the individual lines was expressed as a function of the calibration
error by propagating the equation 
\begin{equation}
V = {\rm c}\frac{\Delta\lambda}{\lambda} 
\end{equation}
and $(\Delta v)^{-2}$ was used as weight.
The calibration error was evaluated as mean shift between the measured position of the night-sky 
emission-lines and their tabulated values and it ranges from 0.38 \AA\ to 0.98 \AA\ for MOS spectra.
These errors, calculated at the H$\alpha$ wavelength, are listed in Table~\ref{radvel}.

\placetable{radvel}

With the new systemic velocity of galaxy 1, the line-of-sight velocity dispersion $\sigma_V$ based 
on
the four brightest group's members (no. 1, 2, 3, and 4) is 65.3 km s$^{-1}$, while the 3D 
velocity dispersion \citep{hi92} is $\sigma_{3D}$ = 95 km s$^{-1}$.
When the tidal dwarf candidate no. 7 is included these values rise to $\sigma_V$ = 149.3 km 
s$^{-1}$
and $\sigma_{3D}$ = 133 km s$^{-1}$, however there are no sufficient information to establish 
whether no. 7 is a self-gravitating object or not.
The dimensionless crossing time, defined by \citet{hi92} as an 
indicator of the dynamical state of a group, is very low: H$_0t_c$ = 0.0067, 
and this should indicate an advanced stage of 
dynamical evolution, although the group does not stringently satisfy the 
correlation between the fraction of late-type galaxies and the crossing time 
found by \citet{hi92}. Actually if we consider the three main galaxies 
1, 2, and 4 (i.e. we exclude the candidate tidal dwarf no. 3), according to the 
results of the photometry (previous Section), the fraction of late-type galaxies is 0.67, so the 
group cannot be considered spiral-poor. However, even if we consider galaxies 1 
and 4 as spirals, no traces of spiral arms are evident from our images, which 
means that if they had spiral arms in the past, they must have lost them as an 
effect of the interaction. This makes it difficult to give an unambiguous 
estimate of the spiral fraction in CG~J1720-67.8.

\subsection{Considerations about the Group's Kinematics}

Kinematics can give important insights on relatively recent interactions 
and/or merger events galaxies went through: such phenomena cause kinematic 
disturbances which are expected to fade within $\sim$ 1 Gyr \citep{ru99}.
Furthermore spiral galaxies' rotation curves can give indications about the 
presence of dark matter halos associated with the galaxies. Therefore the study 
of the kinematics of galaxies in a dense environment such as the one of CG~J1720-67.8,
where strong interactions are currently taking place and could 
have taken place also in the past, would be of fundamental importance.

Unfortunately, considerations about the kinematics of the members of CG~J1720-67.8 
are strongly limited 
by the spatial resolution of the spectra and the small apparent dimensions of 
the group itself. Specifically, velocity curves 
for the member galaxies could not be obtained. Only a few independent
velocity points per galaxy along the slit direction could be measured by 
binning the spectrum in the spatial direction according to the seeing ($\lesssim$ 1$^{\prime\prime}$).
A higher number of points could be obtained only for galaxy 4 along P.A. = 130$^\circ$, 
which do not coincide with either of the galaxy axes. 
Therefore a kinematic study of the individual galaxies and a 
search for anomalies due to gravitational interactions turned out to
be impossible, although it is interesting to note how the radial velocity exhibits
a sort of regular trend between adjacent objects (Fig.~\ref{velcurve}).

Some observations about the kinematics of the tidal arc can be 
done by considering the radial velocities of the knots belonging to it, nos. 3, 
6, 9, 10, 7, 8, assuming that they are actually embedded into the arc and not 
simply projected onto it. As it can be seen from Table~\ref{radvel}, radial velocities in 
the southern part of the arc are lower than the ones in its northern part 
indicating that the southern part is approaching us while the northern part is 
receding. The difference in radial velocity between the southern and the northern tip
of the arc is approximately 300 km s$^{-1}$. 
The minimum value of radial velocity along the arc is found at the position 
of object no. 10.

\placefigure{velcurve}

\subsection{Spectral Classification}

As visible in Figs~\ref{galspec1} and \ref{galspec2}, galaxies 1 and 4 have a blue
continuum and strong emission lines, the latter with some weak absorption features, as well.
The extinction measured from the Balmer decrement in spectrum no. 1 is lower than in spectra
1$\ast$ and 1a (see Table~\ref{galflux}). These measurements suggest a non-homogeneous 
distribution of dust
across galaxy 1, with an apparent concentration in the south-eastern part of it.
The continuum of galaxy 2 is red and there are relatively strong Balmer and metal absorption lines. 
The relative intensity of the blend CaH + H$\epsilon$ with respect to CaK, defined by
\citet{ro84} as an age indicator, indicates that galaxies 2 and 4 have spectra dominated by 
A-type and B-type stellar population, respectively.

For all the objects where H$\beta$ was detectable along with the other emission 
lines, we measured the diagnostic ratios \citep[VO]{vo87} [\ion{O}{3}] $\lambda$ 5007/H$\beta$ 
\emph{vs}
[\ion{O}{1}] $\lambda$ 6300/H$\alpha$, [\ion{N}{2}] $\lambda$ 6583/H$\alpha$, and 
[\ion{S}{2}] $\lambda$ 6716+6731/H$\alpha$ for classification purposes.
The VO diagnostic diagrams shown in Fig.~3 of Paper~I were complemented with the 
additional objects nos. 7, 8, 9, and 10 (Fig.~\ref{VOdia}), thus revealing that all the objects 
belonging to CG~J1720-67.8 have typical emission-line ratios of \ion{H}{2} regions 
or SBNGs \citep{co98}. In the [\ion{S}{2}] diagnostic diagram, the emission knots, TDG candidates
and also galaxy 2 are shifted towards the LINER region. This could indicate the presence of 
shocks: 
such low-ionization lines are strongly enhanced e.g. by shocks produced by outflows of gas,
as occurring in case of stellar winds or supernova remnants \citep{ds95}.
From these diagrams the ionization sources in CG~J1720-67.8 appear to be of thermal nature,
although in the case of galaxy 2 we cannot completely exclude the presence of a low-luminosity
(obscured) active nucleus, due to the somewhat ambiguous values of its emission-line ratios.

\placefigure{VOdia}

In particular the presence of H$\alpha$ emission indicates present-day star formation
in the galaxies and tidal-dwarf candidates, as well as along the tidal arc.
The only object where H$\alpha$ was not detected is no. 11. In its spectrum only 
a \emph{bona fide} [\ion{O}{2}] $\lambda$ 3727 emission line was detected.
Since [\ion{O}{2}] emission can be triggered not only by star formation but also by 
diffuse ionization of the gas, we cannot derive any conclusion about the nature 
of object no. 11 without obtaining a deeper exposure of the object.
The star formation rates of the three galaxies were estimated from their H$\alpha$ intensities
(after correction for internal extinction) adopting the calibration obtained by
\citet{hg86} in the range of mass 0.1 - 100 M$_\sun$:
\begin{equation}
{\rm SFR(M_{\sun}\, yr^{-1})} = 7.07\times 10^{-42} {\rm L_{H\alpha}(erg\, s^{-1})}.
\end{equation}
By considering the area covered by the slit, we deduced the star formation rate densities (SFRD)
per squared parsec. Their values --obtained also from the new spectra-- are given in 
Table~\ref{sfr}.
The strongest starbursts are observed in galaxies 1 and 4 with SFRD reaching $\sim$ 
20$\times$10$^{-8}$
M$_{\sun}$ yr$^{-1}$ pc$^{-2}$, but also the early-type galaxy of the group,
no. 2, shows enhanced star formation. 

\placetable{sfr}

\subsection{Metal Abundances and Photoionization Models}

Metal abundances of the gaseous component could be estimated from emission-line intensity ratios.
Due to the intrinsic weakness of the [\ion{O}{3}] $\lambda$ 4363 line, this was not
observable in our spectra. Such a line is usually visible in the case of particularly low 
abundances or high temperature; however its detection requires a particularly good
signal-to-noise ratio in the spectra.
As a consequence, a direct estimate of the electronic temperature, necessary for 
direct calculations of metal abundances, was not possible.
Therefore we derived a first estimate of the abundances and physical parameters of the
ionized gas by comparison with empirical diagrams \citep{mg91,dtt01}. 
Specifically we compared the R$_{23}$ ratio, defined as $\log(($ [\ion{O}{2}] $\lambda$ 3727 $+$
[\ion{O}{3}] $\lambda$ 4959,5007$)$/H$\beta)$, with the model grid in the plane
[\ion{O}{3}]/[\ion{O}{2}] $vs$ R$_{23}$ in Fig. 10 of \citet{mg91}.
From this comparison, we derived an ionization parameter 
$-$3.5 $\lesssim$ $\log$U $\lesssim$ $-$3 for all the three galaxies.
From this single plot it was not possible to discriminate between the lower and the upper branch of
the R$_{23}$ - $\log({\rm O/H})$ relation. Therefore two possible abundance values were derived
for each galaxy.
Namely $\log({\rm O/H}) \, \sim \, -$3.5 or $\sim \, -$3.8 for galaxy 1,
$\log({\rm O/H}) \, \sim \, -$3.4 or $\sim \, -$4.2 for galaxy 2, and
$\log({\rm O/H}) \, \sim \, -$2.9 or $\sim \, -$4.4 for galaxy 4.

However, if we compare our values of the parameter N2 = $\log($[\ion{N}{2}] $\lambda$ 
6583/H$\alpha )$
with the monotonic relation N2 - 12$+\log({\rm O/H})$ found by \citet[their Fig. 1]{dtt01}
and we refer to the model track with $\log$U = $-$3.0, which is the one better approaching our 
case, 
we find that subsolar abundances (Z $<$ 0.3 Z$_\sun$) are the best suited to our data
for galaxies 1 and 4, while galaxy 2 shows somewhat higher metallicity (Z $<$ 0.5 Z$_\sun$).
Therefore the low metallicity branch of the R$_{23}$ - $\log({\rm O/H})$ relation seems to be the most 
reasonable choice for galaxies 1 and 4, while the high metallicity branch is appropriate to galaxy
2. Thus this first estimate gives the following abundances for the 
gaseous component of the galaxies in CG J1720-67.8:
12$+\log({\rm O/H}) \, \sim$ 8.2, $\sim$ 8.6, and $\sim$ 7.6 for galaxies 1, 2, and 4 respectively.
We used the above values as initial guesses in the calculation of photoionization models
of the three galaxies with the code Cloudy 90.04 \citep{fe96}.
In accordance with the \ion{H}{2} nature of the galaxies established above (Fig.~\ref{VOdia}), 
we assumed as ionizing source a thermal continuum. Specifically we adopted a Mihalas continuum
with T = 40000 K. The ionization parameter estimated from the diagram of \citet{mg91} and
the electronic density estimated from the [\ion{S}{2}] $\lambda$ 6716/[\ion{S}{2}] $\lambda$ 6731
ratio assuming an electronic temperature T$_{\rm e}$ $\sim$ 10$^4$ K were used as initial 
parameters. The hydrogen density and the ionization parameter were allowed to vary to reach the best-fit
to the emission line intensities relative to H$\beta$. Several models were calculated with slightly
modified metal abundances until a reasonable fit to the line ratios was obtained.
The parameters and metal abundances of the best-fit models are given in Table~\ref{modelparam}, 
and
the observed and modeled line ratios are listed in Table~\ref{photmodel}.
The galaxy with lowest metal abundances is no. 4, having Z $\sim$ 0.1 Z${_\sun}$, while the 
highest 
abundances, Z $\lesssim$ 0.5 Z${_\sun}$, are found for galaxy 2.
The values of the ionization parameter are in agreement with the above first estimates:
$\log$U $\sim$ $-$3.2 for galaxy 1 and $\log$U $\sim$ $-$3.6 for galaxies 2 and 4.

\placetable{modelparam}
\placetable{photmodel}

\subsection{FIR Emission and Dust Content: Not a Puzzle Anymore?}

The amount of internal extinction determined from the Balmer decrement 
in the galaxies and in the northern part of the tidal arc of CG~J1720-67.8 
is considerable (Tables~\ref{galflux} and \ref{tdgflux}). This is actually not
a surprise, given the starburst nature of these objects and the
high concentration of dust which is usually accompanying episodes of
intense star formation. The dust seems to be concentrated mainly in the 
northern part of the group.

As a consequence of the enhanced star formation in the
group's objects and of the noticeable amount of interstellar dust, a
high FIR luminosity is expected due to re-radiation by the dust of 
the UV light emitted by young hot stars.
Nevertheless CG~J1720-67.8 has not been cataloged as
IRAS source. This could be due to the fact that the group -- having
Galactic coordinates $l$ $\sim$ 324\degr, 
$b$ $\sim$ -16\degr -- is located in a region of the sky
dominated at 100 $\mu$m by ``infrared cirrus'', whose
presence makes it difficult to identify real point sources and might
corrupt the measurement of their fluxes. 

However an inspection of IRAS raw data at the position of the group
revealed the presence of emission apparently peaked at 60 $\mu$m and
a hint of emission at 100 $\mu$m, although difficult to disentangle
from the extended structures visible in this band.
Low resolution co-added maps in these two bands, centered on the position 
of CG~J1720-67.8 with a field size of 32$^{\prime}\times$32$^{\prime}$ 
were extracted from position- and flux-calibrated raw IRAS data --
obtained from the IRAS server at the Space Research Organization
Netherlands SRON -- by use of the
program GIPSY\footnote{Groningen Image Processing SYstem, available at: 
http://www.astro.rug.nl/$\sim$gipsy/}.
In order to obtain a higher (by a factor of $\sim$ 5) spatial resolution, 
approaching the diffraction limit of the satellite
($\sim$ 1$^{\prime}$ at 60 $\mu$m and $\sim$ 1$^{\prime}$.7 at 100 $\mu$m), 
the HIRAS program, which drives the MEMSYS5 maximum entropy imaging algorithm
\citep{bo94} was applied to the data. The resulting FIR ``images'' 
and the corresponding contour maps overlapped to the Digitized Sky
Survey (DSS) image of the galaxy group are shown in Fig.~\ref{IRASmaxent}.
Even with such a resolution -- assuming that the observed emission is
really associated with the galaxy group -- it is not possible to
distinguish FIR emission 
stemming from individual group's members, due to the small size of the 
galaxy group itself ($\sim$ 30$^{\prime\prime}$ in diameter), 
which is well below the
diffraction limit of IRAS. Therefore it is not clear whether the detected FIR
emission is associated with the whole group or only with one or more of its components.
 
\placefigure{IRASmaxent}

The flux density of the source in the two FIR bands was estimated
with the GIPSY task FLUX inside three different polygonal apertures defined
``by hand'' around the source just outside the faintest border of the
source visible in the images after the application of the maximum
entropy algorithm.
The mean of the three measured values and the associated $rms$ were taken as 
the flux density measurement and its error. The background value, obtained
as mean value of the mode inside three different boxes free of sources in 
the same field and multiplied by the number of pixels in the polygonal apertures,
was subtracted from the measured fluxes.
The resulting flux densities at 60 and 100 $\mu$m are F$_{60}$ = 1.1
$\pm$ 0.2 Jy
and F$_{100}$ = 3.6 $\pm$ 0.7 Jy, from which the total FIR flux
between 42.5 and 122.5 $\mu$m could 
be derived following \citet{he85}:
\begin{equation}
\rm FIR = 1.26\times 10^{-11} (2.58 F_{60} + F_{100})
= 8.1\times10^{-11} ~erg ~cm^{-2} ~s^{-1}
\end{equation} 
yielding a FIR luminosity L$_{\rm FIR}$ = 3.16 $\times$
10$^{44}$ erg s$^{-1}$. Therefore, following \citet{hu86}, 
the total star formation rate of the group can be evaluated as:
SFR$_{\rm FIR}$ = 1.34$\times$10$^{-43}$L$_{\rm FIR}$ = 42.3 M$_{\odot}$ yr$^{-1}$, 
a value a factor of a few  higher than 
the one which could be deduced by summing up the SFR of all the group's 
objects as measured from their H$\alpha$ luminosities.

By assuming a single temperature component and a $\lambda^{-1}$
emissivity law, also estimates of the
temperature and mass of the ``warm'' dust -- having a temperature T
$\geq$ 25 K -- could be derived with the following
expressions \citep{yo89,br98}:
\begin{equation}
\rm T_{dust} \simeq 49\, (F_{60}/F_{100})^{0.4} \, \, \, K
\end{equation}
\begin{equation}
\rm M_{dust} = 4.78\, F_{100}\, D^2 [\exp (143.88/T_{dust})-1]
\, \, \,  M_{\odot}
\end{equation}
where the distance D is expressed in Mpc and $\rm F_{60}$
and $\rm F_{100}$ are in Jy. These calculations yielded $\rm T_{dust}$ $\sim$ 30 K
and $\rm M_{dust}$ $\sim$ 6.2$\times$10$^7$ M$_{\odot}$.

\subsection{System Mass Estimate}

The traditional methods of mass estimate of a galaxy system are all based on the assumption that
the system has reached the virial equilibrium.
Numerical simulations (Perea, Del Olmo, \& Moles 1990) have shown that such a condition is
reached after about three crossing times. Before that time, mass estimators assuming 
virial equilibrium might lead to considerable errors in mass.
Despite the short crossing time we estimated for CG~J1720-67.8, we cannot state that this group is
already virialized. The presence of an asymmetric optical halo, the strong signs of
recent interactions, the prominence of (typically short-lived) tidal 
features, the interaction induced bursts of star formation, which can be dated $\sim$ 40 -- 180 Myr 
according to evolutionary synthesis models \citep{tefva02}, are all elements that 
suggest CG~J1720-67.8 is most probably not yet in virial equilibrium. Furthermore, only projected 
separation among galaxies and
radial velocities are known, and the lack of three-dimensional information
could lead to large errors in the mass of small groups.

Even so, we attempted to obtain an indication of the system mass by applying the four 
mass estimators described in Heisler, Tremaine, \& Bahcall (1985): virial mass (${\cal M}_V$), 
projected mass (${\cal M}_P$), median mass (${\cal M}_M$), and average mass (${\cal M}_A$).
All of these estimators, except for ${\cal M}_M$, are sensitive to the presence of interlopers,
but in our case the small radial velocity difference among the group's members and the 
evidence of interactions assure the absence of interlopers.

Although \citet{htb85} claimed that ${\cal M}_V$ calculated under the assumption of equal mass
bodies is accurate even in case of galaxies of different mass, \citet{pe90} found that the 
absence of virial equilibrium can result in an overestimate as large as a factor of two of
${\cal M}_V$ when a weighting over the mass of the galaxies 
(or alternatively over their luminosity, by assuming an adequate M/L ratio) is not performed.
Nevertheless, given the lack of adequate kinematic information, we are not able
to estimate the masses of the members of CG~J1720-67.8.
Albeit estimates of the masses of the three galaxies have been obtained from the
comparison of the observed luminosities with the best-fit chemically consistent evolutionary
synthesis models taking into account the contribution of the starbursts to the galaxy luminosities 
\citep{tefva02}, M/L ratios for the TDG candidates cannot be reliably estimated.
Up to date the only example of CG where the masses of candidate TDGs have been measured is
Stephan's Quintet. Its candidate TDGs have been found to have a median M/L$_{\rm B}$ ratio of 7 
\citep{mo01}, but with a very large scatter. We find that assuming such a value for our TDG candidates
would not be meaningful. Therefore we calculated only the non-weighted virial mass of the group.
The application of the four mass estimators cited above gave the following values 
(expressed in M$_{\sun}$ units):
${\cal M}_V$ $\simeq$ 7.0$\times$10$^{11}$, 
${\cal M}_P$ $\simeq$ 2.6$\times$10$^{11}$,
${\cal M}_M$ $\simeq$ 3.3$\times$10$^{11}$, and
${\cal M}_A$ $\simeq$ 8.1$\times$10$^{11}$.

We note that the projected mass, less affected by the mass spectrum, is a factor of 2.6 lower
than the virial mass.

The lack of sufficient information on the mass of the individual group members (in particular TDG candidates)
as well as on the gaseous mass of the group (no radio nor X-ray observations are available for
CG~J1720-67.8) prevent us from giving an estimate of the dark matter content of the group.

\section{FURTHER DISCUSSION AND CONCLUSIONS}

We have presented a detailed photometric and spectroscopic study of CG~J1720-67.8,
a rare example of compact group that allows an observation of the very process whereby 
galaxies are stripping themselves.
We have found that the star formation activity is not confined to the three main galaxies 
of CG~J1720-67.8, but is spread over the whole group, including its extended tidal tail.
We have concentrated our attention mainly on the galaxies (objects no. 2, 4, and 1 in 
order of decreasing luminosity), and we have attempted a quantitative morphological 
classification
and a quantification of the interaction effects on their morphology, by means of bidimensional
fitting and bulge-disk decompositions.
From our investigation the presence of morphological distortions in all the three galaxies
appeared clearly. Despite the application of quantitative methods, an unambiguous morphological
classification turned out to be difficult, because these galaxies are already about to loose their
identity as a consequence of the interaction processes in which they are involved. 

Nonetheless,
we conclude that galaxy no. 2, is probably an S0, with a B/T ratio typical of a bulge-dominated
system and a bulge which can be well reproduced by a de Vaucouleurs law, once the seeing effects are
taken into account. The color analysis of this galaxy shows that it is considerably reddened by
dust. Actually its optical spectrum exhibits a red continuum, although its absorption lines
indicate that it is dominated by a relatively young (A-type) stellar population, and the presence
of emission lines suggests recent star formation activity most likely concentrated in the central
part of the galaxy. Emission line ratios and photoionization models indicate that, with its
Z $\sim$ 0.5 Z$_{\sun}$, this is the galaxy with the highest metallicity inside the group.

Galaxy 4 appears to be a disk-dominated system with internal structures reminiscent of knotty
spiral arms apparently connected with the extended arc-shaped tidal feature of the group.
Its spectrum indicates the starburst nature of the galaxy.
This is probably an Sc galaxy, whose spiral structure has been already destroyed by the
interaction.

Galaxy 1 shows blue colors and the most intense star formation activity inside the
group. Its B/T ratio is
typical of a bulge-dominated system. Although its surface brightness distribution can be fitted
with a de Vaucouleurs bulge plus an exponential disk, a fit with a S\'ersic bulge would give
a S\'ersic index $n\, \simeq$ 1, i.e. an exponential bulge, similar to that found in late-type
spirals \citep{as94,a98}. The galaxy is somewhat elongated towards galaxy 2, to which it seems to be
connected by a bridge of matter. It is not yet clear whether this galaxy is an Scd system, whose
outermost layers have been completely stripped away to leave a dominating exponential bulge, 
or it is a small disk system whose B/T ratio is increased as an effect of a central starburst
faking a bulge structure.

The star formation activity of the group is accompanied by a patchy distribution of dust,
with a concentration in the north-eastern part of it.
However, once the star formation rates estimated from the H$\alpha$ emission are considered,
the expected FIR flux for this $\sim$ 180 Mpc distant object is only slightly above the IRAS
detection limit. Furthermore the Galactic latitude of the group (b $\sim \, -$16) locates it in a
region of the sky particularly affected by Galactic cirrus in the FIR. 
This contamination by Galactic cirrus can explain why the
group is not cataloged as an IRAS source, although our analysis of IRAS raw data revealed a
measurable source at the coordinates of CG~J1720-67.8, whose flux is consistent with the 
measured SFR. 

The virial mass of the group estimated by taking into account all the group members
is $\sim$ 7$\times$10$^{11}$ M$_{\odot}$, a factor of 2.6 higher than the
projected mass. The virial mass could actually be overestimated since the mass spectrum of
the group was not taken into account. A comparison with the estimated total mass of the group with 
the luminous mass is not possible with the data currently available -- in particular the mass of
TDG candidates as well as the total gas content of the group, cannot be estimated. 
Therefore no conclusions can be drawn about the dark matter content of the group.
Since the dark matter content and distribution play an important role in the evolution of groups
(see e.g. numerical simulations by \citet{amb97}), further investigations are needed.

\acknowledgments

We are grateful to the anonymous referee, whose comments and suggestions have allowed us to
significantly improve this paper. Thanks to the suggestions of the scientific editor L. S. Sparke
and to the new images provided by L. Infante, we were able to better assess the reliability of 
our optical photometry.
A part of this work was supported by the Austrian Science Fund (FWF) under project 
no. P15065. RW is grateful to the Austrian ``Bundesministerium f\"ur Wissenschaft and Verkehr"
for travel support.

\appendix
\section{Reliability of Optical Photometry}

Since there are no (non-saturated) stars in the galaxy group's frames with 
cataloged B,V, and R magnitudes available, it was impossible to use some of the stars 
in these frames
either for performing the photometric calibration or for testing in a direct 
way the one we adopted, thus an alternative way to verify the reliability of the
optical photometry had to be followed.

Due to the importance of the use of color indices for the purposes of our 
analysis, we tried to reveal effects of variations in the atmospheric conditions 
on the measured colors through photometric measurements of the stars contained 
in the same frames of CG~J1720-67.8. For this purpose we measured 65 
non-saturated stars visible in those B, V, and R images, which we used for the photometry
of CG~J1720-67.8.
The magnitudes of the stars were measured with the IRAF task PHOT
inside apertures of 4$^{\prime\prime}$ in radius. Errors were calculated 
according to the following expression:
\begin{equation}
\Delta m = \frac{1.0857}{F} \left(\frac{F}{gain} + A \sigma^2 + A^2 \frac{\sigma^2}{nsky}\right)^{1/2}
\end{equation}
where $F$ is the flux in ADU inside the aperture of area $A$, 
$nsky$ is the number of pixels used for the sky determination, $\sigma$ is the 
standard deviation and the CCD gain is expressed in e$^{-}$ ADU$^{-1}$.

Specifically, we compared the distribution of such field stars in color-color 
diagrams with the distribution of colors from synthetic spectra calculated by 
\citet{bcp98} from the \citet{ku95} grids of model 
atmospheres. Table 1 of Bessel et al. (1998), containing the synthetic colors, 
is available via anonymous ftp from cdsarc.u-strasbg.fr.
The resulting color-color diagrams in Fig.~\ref{star2color} show agreement within the 
error measurements between most of the stars and the distribution of the 
synthetic colors. In fact, except for a few points, most of the stars
agree with the synthetic tracks better than 0.05 mag. The 6 deviant points all lie
above the synthetic track in both color-color diagrams, indicating that in the R image they
appear brighter than expected with respect to their B and V magnitudes.
After an inspection of the R image we found that one of these points is a star contaminated
by an edge-on background galaxy, which becomes visible only in the deep R image.
The other peculiar points correspond to stars lying in crowded regions and contaminated by 
the halos of saturated stars.
Note that a systematic effect in one of the bands would cause 
a uniform shift of all the observational points above or below the track defined 
by synthetic points. No such effects can be observed in Fig.~\ref{star2color}, therefore if a 
systematic effect is present, it must be affecting all three bands by the 
same amount and has no consequences in the optical colors. Although we consider the
possibility as unlikely that non-photometric conditions of the sky have generated such an effect, 
because the
images in the three bands were all taken in different nights, we performed a further 
test to assess the reliability of the optical absolute magnitudes. For this purpose we attempted
an alternative procedure for calibrating our images.

New BVR short exposures (240 s in B, 120 s in V and R) of the galaxy group were obtained at 
the ESO 3.6 m telescope with EFOSC2 and kindly provided by L. Infante. 
After the application of the standard reduction procedure to 
these new images, we proceeded as follows:

\begin{enumerate}
\item We measured on the new BVR images aperture magnitudes of 10 bright stars, for which $B_{\rm J}$ and 
$R_{\rm F}$\footnote{The correct nomenclature for these filters would be J and F, but we
adopt here the alternative one $B_{\rm J}$ and $R_{\rm F}$ to avoid confusion with 
the NIR band J.} magnitudes are given in the Guide Star Catalog 
(GSC-2\footnote{http://www-gsss.stsci.edu/gsc/gsc2/GSC2home.htm}).
We used these stars as standard to calibrate the three images in the $B_{\rm J}R_{\rm F}$ system 
by means of the following transformation equations:

\begin{equation}
B_{\rm J} = b + k1 + k2(B_{\rm J} - R_{\rm F})
\end{equation}
\begin{equation}
B_{\rm J} = v + k3 + k4(B_{\rm J} - R_{\rm F})
\end{equation}
\begin{equation}
R_{\rm F} = r + k5 + k6(B_{\rm J} - R_{\rm F}),
\end{equation}

where $b, \, v$, and $r$ are the instrumental magnitudes measured in the B, V, and R
images and kn (n = 1, ..., 6) are the zero points and color terms to be determined through 
the fit to the standard stars. 

\item Out of the 65 stars already used for the tests on optical colors, we identified 24 with known
$B_{\rm J}$ and $R_{\rm F}$ magnitudes in the GSC-2. We measured their aperture magnitudes
in the new BVR images and applied the above transformation equations.
This operation gave us measured $B_{\rm J}$ and $R_{\rm F}$ magnitudes to be compared with 
the cataloged ones. We found a good agreement between cataloged magnitudes and measured magnitudes
obtained from the B and R frames, while the magnitudes obtained from the V frame show significant
deviations (0.2 mag on average) from the catalogued $B_{\rm J}$ magnitude.
In the following we use the cataloged magnitudes of these stars, but the measurements and comparison
we have done here indicate that the V filter offers a poor match to the 
$B_{\rm J}R_{\rm F}$ system.

\item We selected a set of 15 Landolt standard stars approximately matching in colors the stars
in our frames and having $B_{\rm J}$ and $R_{\rm F}$ magnitudes available from the GSC-2.
We used these stars to obtain the transformation equations from the $B_{\rm J}R_{\rm F}$ system 
to the Johnson-Cousins BVR system: 

\begin{equation}
B = B_{\rm J} + c_{\rm B}(B_{\rm J}-R_{\rm F})
\label{systransfB}
\end{equation}
\begin{equation}
V = B_{\rm J} + c_{\rm V}(B_{\rm J}-R_{\rm F})
\label{systransfV}
\end{equation}
\begin{equation}
R = R_{\rm F} + c_{\rm R}(B_{\rm J}-R_{\rm F}),
\label{systransfR}
\end{equation}

which yielded the following color terms: $c_{\rm B}$ = 0.071$\pm$0.026, $c_{\rm V}$ =
-0.574$\pm$0.015, and $c_{\rm R}$ = 0.030$\pm$0.015.
When the fits are evaluated on the standard stars themselves (IRAF task PHOTCAL.EVALFIT) 
they give median residuals with respect to the cataloged values of -0.019 in B, -0.024 in V
and 0.003 in R, with large scatters (up to 0.2 mag) in B and V.
Therefore the photometric system transformation will be a significant source of error especially in
the calibration of B and V bands.

\item We applied the transformations (A5), (A6), and (A7) to the set of 24 faint
stars with known magnitudes selected in the field of view of the galaxy group (see point 2 above).
Thus we obtained a set of non-saturated BVR stars to be used as standard for the calibration 
of our original BVR images. We used transformation equations analogous to (1), (2), and (3)
(Section 3.2), only omitting the term concerning the atmospheric extinction correction, 
which is now implicitly included in the zero point of the photometric system.
The newly found zero points, B$^{\prime}_0$ = 25.63$\pm$0.07, V$^{\prime}_0$ = 25.85$\pm$0.03, and 
R$^{\prime}_0$ = 25.92$\pm$0.04, agree within the errors with 
the zero points we have used to calibrate the optical images of the galaxy group (once the
extinction correction has been included, see Section 3.2);
in fact the differences between the zero points we used and the new ones are 
-0.03, -0.09, and -0.008 in B, V, and R,
respectively. Note that the errors reported here are the ones given by the IRAF task 
PHOTCAL.FITPARAM 
by performing a uniformly weighted fit, therefore they do not account for the errors due to the 
transformation between the $B_{\rm J}$$R_{\rm F}$ and the Johnson-Cousins BVR system. Furthermore
the V band is badly calibrated with $B_{\rm J}$ magnitudes as seen in point 2 above.
The errors given by FITPARAM for the color terms are comparable to the values themselves, and
again this is an underestimate of the actual error. 

\item We finally applied the new calibration to all the 65 stars previously used for the optical 
color reliability test and we built again color-color diagrams analogous to those in
Fig.~\ref{star2color}.
The result is shown in Fig.~\ref{newstar2color}. The match of the observational points to the 
synthesized tracks
is considerably poorer than the one in Fig.~\ref{star2color}, especially for redder colors. 
This is probably
a consequence of the numerous sources of errors in the transformations applied to obtain the
new photometric calibration.

\end{enumerate}

We conclude that an alternative calibration based on cataloged stars with known $B_{\rm J}$
and $R_{\rm F}$ magnitudes would be still consistent with the one we adopted, but would be affected
by larger errors.

\placefigure{star2color}
\placefigure{newstar2color}

\clearpage
\onecolumn
\begin{figure}
\figcaption{A blow-up of CG~J1720-67.8 from the average of three 600 s R-band exposures 
obtained at the ESO 3.6 m telescope.
The positions at which optical spectra were acquired are labeled. The three main galaxies
of the group are nos. 1, 2, and 4. Nos. 3 (or 3+9) and 7 are the most promising tidal dwarf 
candidates. \label{Rgroup}}
\end{figure}

\begin{figure}
\figcaption{Sequence of optical and NIR images of CG~J1720-67.8 (in logarithmic color scale),
from the B band through K$\rm _s$. The field size is $\sim$ 
1.2$^{\prime}\times$1.0$^{\prime}$. \label{allbands}}
\end{figure}

\begin{figure}
\figcaption{Optical (top) and NIR (bottom) color maps of CG~J1720-67.8. The color scale of the
images is selected in a way such that physically
bluer structures appear blue in the images and redder structures appear
red. The positions of all objects are labeled according to Fig.~\ref{Rgroup}.\label{allcolors}}
\end{figure}

\begin{figure}
\figcaption{Left: Residual B-band image after the subtraction of galaxy 2, whose center position
is indicated with a cross. Right: Residual B-band image after the subtraction of
galaxies 1 and 4, whose center positions are marked with crosses.
A comparison between the two images shows that there is significant overlapping between
the regions occupied by the three galaxies.\label{bmodel}}
\end{figure}

\begin{figure}
\figcaption{Residual image (R band) of CG~J1720-67.8 after the subtraction of the best-fit models
of galaxies 1, 2, and 4 obtained with the package GIM2D. Labels 1, 2, and 4 indicate the
centers of the subtracted galaxies. \label{gim2dres}}
\end{figure}

\begin{figure}
\epsscale{.8}
\plotone{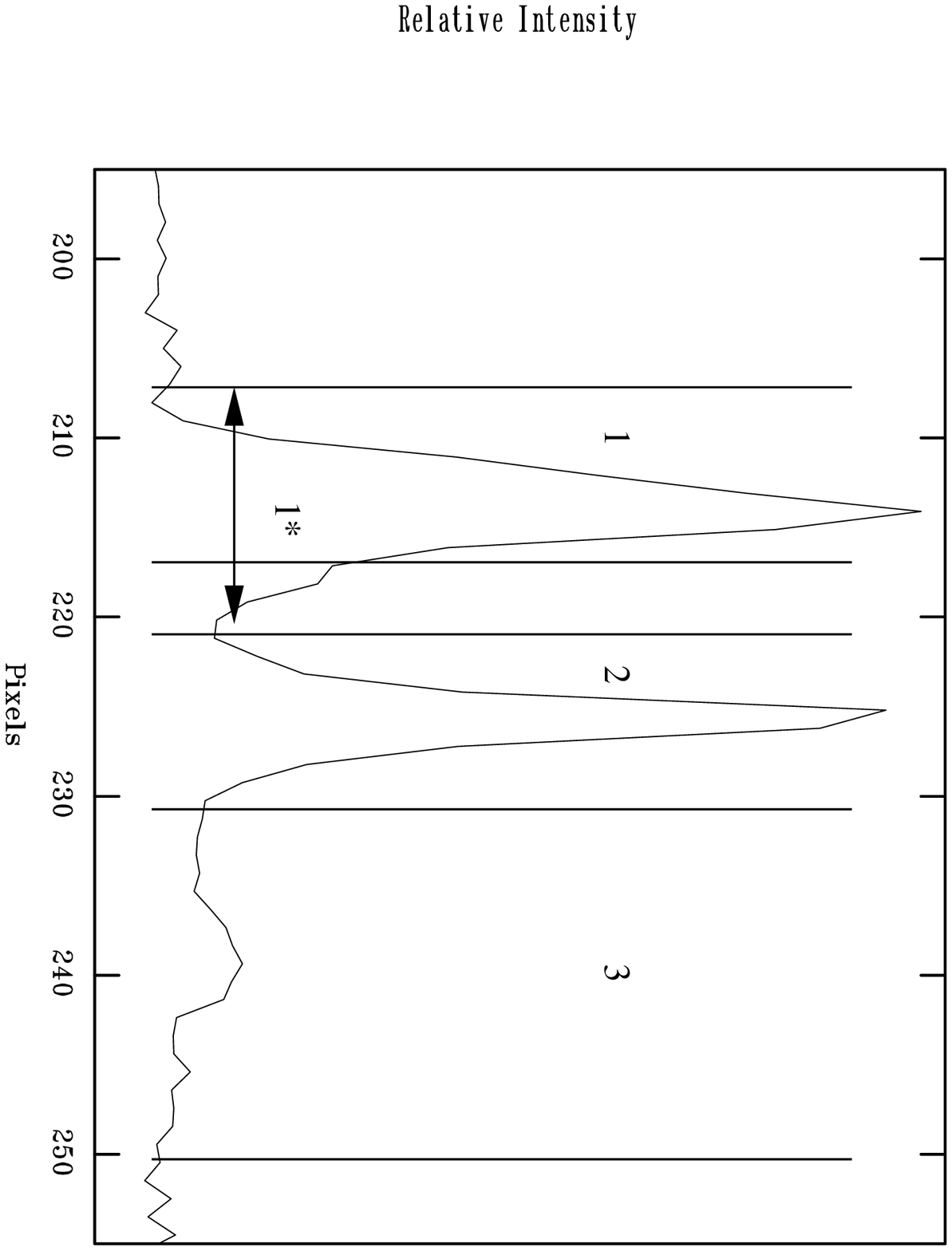}
\caption{H$\alpha$ profile along the slit, used for the identification and extraction of
objects 1, 2, and 3 from the long-slit spectrum obtained at the Du Pont telescope in Las
Campanas. The portions of the spectrum occupied by the three objects are marked with vertical
lines. The profile of galaxy 1 exhibits a secondary peak. 
We extracted the most intense peak as
spectrum no. 1 and the two peaks together as spectrum 1$\ast$.\label{Haprof}}
\end{figure}

\clearpage

\begin{figure}
\epsscale{1.0}
\figcaption{Spectra of galaxy 1. The upper two are long-slit ones at P.A. = 130\degr, while
the lower two are MOS spectra at P.A. = 90\degr. Spectrum 1b confirms that the bright knot
adjacent to the nucleus of galaxy 1 is actually a Galactic star projected onto the galaxy.
 \label{galspec1}}
\end{figure}

\begin{figure}
\plotone{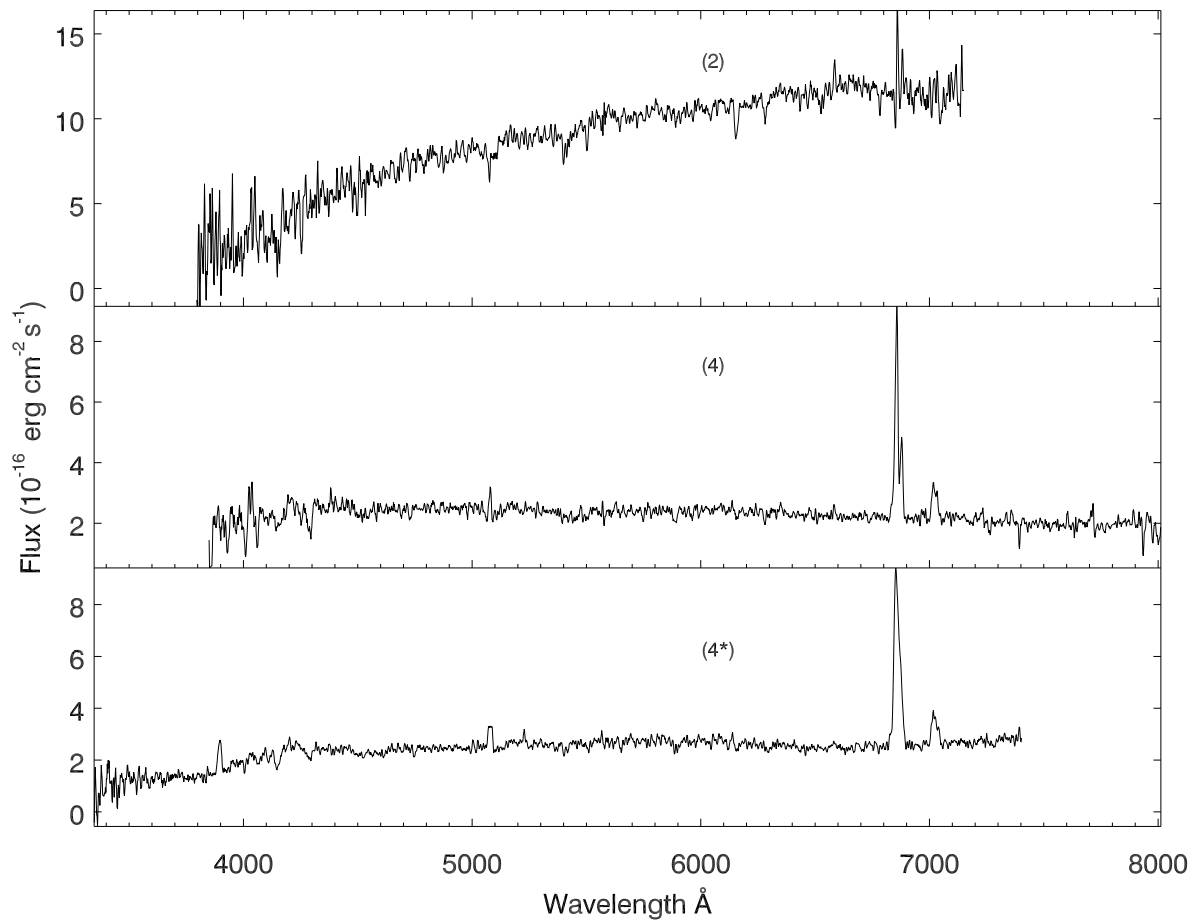}
\caption{Long-slit spectra of galaxies 2 (up) and 4 (middle) at P.A. = 130\degr, and
MOS spectrum of galaxy 4 (bottom, labeled 4$\ast$) at P.A. = 90\degr. \label{galspec2}}
\end{figure}


\begin{figure}
\plotone{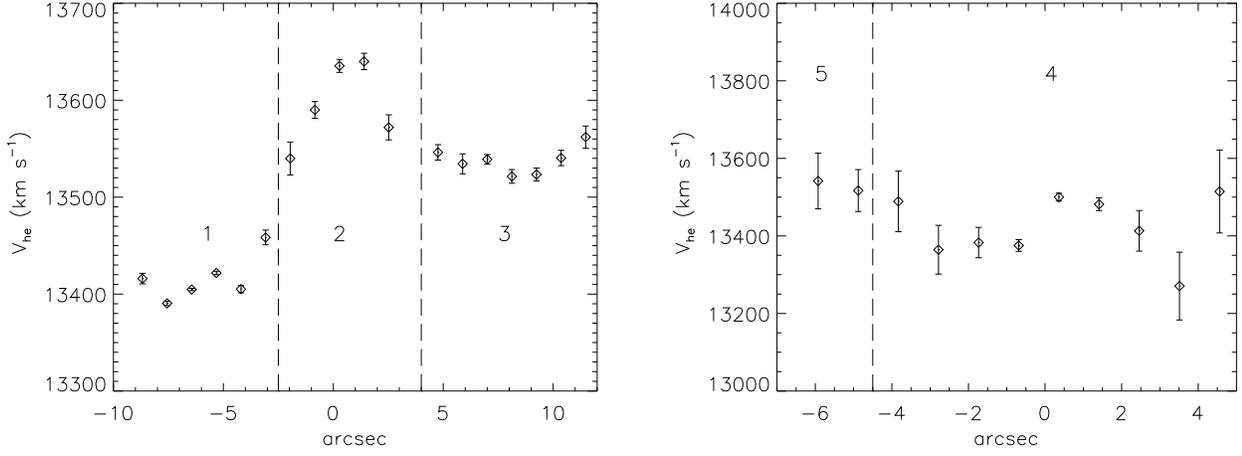}
\caption{Trend of heliocentric radial velocity for objects 
nos. 1, 2, and 3 (left) and nos. 4
and 5 (right) obtained from the measurement of the H$\alpha$ position 
along the slit (P.A. = 130$^\circ$). On the x axis the positions in
arcsec relative to the position of the peak of galaxy 2 (left) and of
galaxy 4 (right) are given with positive values eastward. Dashed
vertical lines mark the separation between adjacent objects. \label{velcurve}}
\end{figure}

\begin{figure}
\epsscale{.35}
\plotone{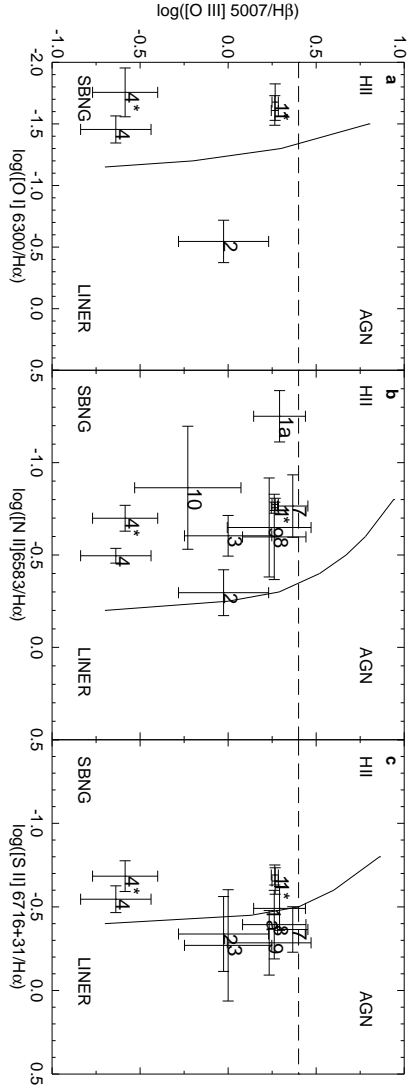}
\caption{VO diagnostic diagrams \citep{vo87} for all the galaxies, knots, and
TDG candidates in CG~J1720-67.8. The horizontal
dashed line additionally discriminates between HII-like regions and
starburst nucleus galaxies (SBNG) on the left side, and between AGN
and low ionization nuclear emission regions (LINER) on the right side.\label{VOdia}}
\end{figure}

\begin{figure}
\figcaption{Top -- 60 and 100 $\mu$m IRAS maps of a
$32^{\prime}\times32^{\prime}$ field centered at the coordinates of CG~J1720-67.8 
after the resolution improvement obtained by applying the
maximum entropy algorithm. 
Intensities are increasing from darker to brighter colors.
Bottom -- Contour maps of the $8^{\prime}.5\times8^{\prime}.5$ central 
region of the above images overimposed to the DSS image of the group.
The displayed contour levels are at 1 and 2 $\sigma$ (dashed line) and 
3, 4, 5, 6, and 10 $\sigma$ (solid line) above the background. \label{IRASmaxent}}
\end{figure}

\begin{figure}
\plotone{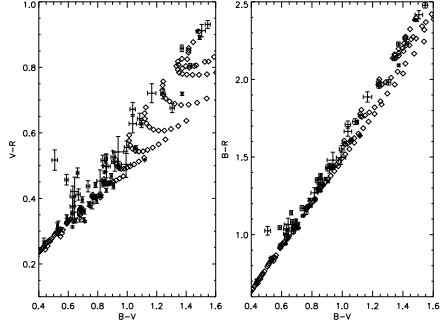}
\caption{Optical color-color diagrams where observed stars belonging to the 
same frames of CG~J1720-67.8 (points with associated error bars) are compared with the 
distribution of the 
synthetic colors (diamonds) given in Table 1 of Bessel et al. (1998).\label{star2color}}
\end{figure}

\begin{figure}
\plotone{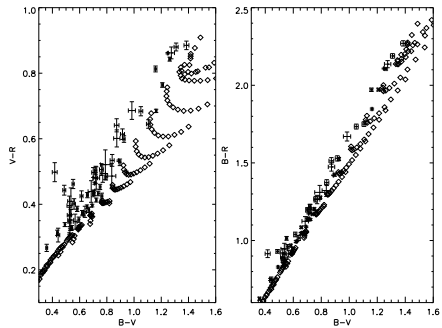}
\caption{Same as Fig.~\ref{star2color}, but the observational points have been 
calibrated in an alternative way, by use of stars of known $B_{\rm J}$ and $R_{\rm F}$ magnitudes,
appropriately converted into the Johnson-Cousins BVR system. \label{newstar2color}}
\end{figure}

\clearpage

\begin{deluxetable}{llcccccccc}
\tabletypesize{\scriptsize}
\tablecaption{Observational Details \label{obslog}}
\tablewidth{0pt}
\rotate
\tablehead{
\colhead{Telescope} & \colhead{Instrument} & \colhead{Filter/Grism/Mask} & 
\colhead{Date} & \colhead{Exp. Time} & \colhead{Seeing} & \colhead{Spatial Scale} &
\colhead{Spec. Res.}\tablenotemark{a} & \colhead{Disp.} & \colhead{Spec. Range}\tablenotemark{b}\\
 & & & & \colhead{(s)}& \colhead{($^{\prime\prime}$)} & \colhead{($^{\prime\prime}$ px$^{-1}$)}
& \colhead{(\AA)} & \colhead{(\AA px$^{-1}$)} & \colhead{(\AA)}}
\startdata
ESO 3.6m & EFOSC1 & Johnson I\tablenotemark{c} & 1995 Jun 28& 900 & 1.6& 0.61 & & & \\
ESO-MPIA 2.2m & EFOSC2 & GR\#6 &1997 May 4 & 1200 & $\sim$ 1 & 0.262 & $\sim$12 & 
2.052 & 3850 -- 7950\\
Du Pont 2.5m  &Mod. Spec. &Grat. 600 l/mm &1998 Feb 13 &1500 &$<$ 1 & 0.56 & $\sim$4&2.005 
& 3770 -- 7180\\
Du Pont 2.5m & CCD Tek\#5 & Johnson V & 1998 Feb 15 & 900 & 1.7 & 0.259 & & & \\ 
ESO 3.6m & EFOSC2 &Johnson  R &1999 Jun 7 & 480 & 1.3 & 0.157& & & \\
ESO 3.6m & EFOSC2 &Johnson  B &1999 Jun 7 & 900 &1.5 & 0.157& & & \\
ESO 3.6m & EFOSC2 &Johnson  R &1999 Jun 15 &3 $\times$ 600s &1.5 &0.157 & & & \\ 
ESO 3.6m & EFOSC2 &Johnson  B &1999 Jun 15 - 16& 2 $\times$ 1800s& 1.6& 0.157& & & \\
ESO 3.6m & EFOSC2 &Johnson  V &1999 Jun 16& 600 & 1.8& 0.157& & & \\ 
ESO 3.6m & EFOSC2-MOS &GR\#11 M\#1 & 1999 Jun 16& 1800 &$\geq$ 1.4 & 0.157&$\sim$19 
& 2.05 & 3350 -- 7560 \\
ESO 3.6m & EFOSC2-MOS &GR\#11 M\#2 & 1999 Jun 16& 1800 &$\geq$ 1.4 & 0.157&$\sim$19 
& 2.05 & 3350 -- 7560 \\ 
ESO 3.6m & EFOSC2-MOS &GR\#11 M\#3 & 1999 Jun 16& 1800 &$\geq$ 1.4 & 0.157&$\sim$19 
& 2.05 & 3350 -- 7560 \\
ESO 3.6m & EFOSC2-MOS &GR\#11 M\#4 & 1999 Jun 16& 1800 &$\geq$ 1.4 & 0.157&$\sim$19 
& 2.05 & 3350 -- 7560 \\ 
ESO 3.6m & EFOSC2-MOS &GR\#11 M\#5 & 1999 Jun 16& 1800 &$\geq$ 1.4 & 0.157&$\sim$19 
& 2.05 & 3350 -- 7560 \\
Swope 1m &IRC-NICMOS3  & J    & 2000 Apr 22& 90 $\times$ 20s& 1.4 & 0.1498\tablenotemark{d} & & &\\ 
Swope 1m &IRC-NICMOS3  & H    & 2000 Apr 22& 40 $\times$ 50s& 1.5 & 0.1498\tablenotemark{d} & & &\\
Swope 1m &IRC-NICMOS3  & K$\rm _s$& 2000 Apr 22& 50 $\times$ 50s& 1.4 & 0.1498\tablenotemark{d} & & &\\ 
\enddata
\tablenotetext{a}{Spectral resolution at $\lambda$ $\sim$ 6800 \AA, i.e. at the redshifted H$\alpha$
wavelength.}
\tablenotetext{b}{Spectral ranges given for MOS spectra are only indicative, 
as the actual spectral range was changing with the position of the slitlets on the field.}
\tablenotetext{c}{The I-band image could not be calibrated.}
\tablenotetext{d}{Spatial resolution after subpixeling the original resolution of 0.599 "/pix
(0.599/4).}
\end{deluxetable}

\clearpage

\begin{deluxetable}{cccccccc}
\tabletypesize{\footnotesize}
\tablecaption{Fluxes and Intensities of the Galaxies in CG~J1720-67.8 \label{galflux}}
\tablewidth{0pt}
\tablehead{
\colhead{Line} & \colhead{Spec. 1$\ast$} & \colhead{Spec. 1} & \colhead{Spec. 1a} &
\colhead{Spec. 2} & \colhead{Spec. 4} & \colhead{Spec. 4$\ast$}}
\startdata
$[$OII$]$ $\lambda$3727  & 2.38$\pm$0.36 & 1.79$\pm$0.29
&2.05$\pm$0.70 & 3.86$\pm$2.28 & ... &0.92$\pm$0.30 \\
                         & {\bf  3.26}   & {\bf  2.28}   & {\bf  3.70}
& {\bf  6.23} &{\bf ... } & {\bf  1.70} \\
H$\delta$                & 0.39$\pm$0.18 & 0.42$\pm$0.14 & ...
& ... & ... & ...  \\
                         & {\bf  0.49}   & {\bf  0.49}   & {\bf ... }
& {\bf ... } & {\bf ... } &  {\bf ... }\\
H$\gamma$                & 0.42$\pm$0.06 & 0.42$\pm$0.05 &
0.39$\pm$0.24 & ... & ... & ...  \\
                         & {\bf 0.49}    & {\bf  0.47}   & {\bf  0.53}
& {\bf ... } & {\bf ... } &  {\bf ... }\\
$[$OIII$]$ $\lambda$4959 & 0.65$\pm$0.05 & 0.63$\pm$0.04 &
1.02$\pm$0.47 & 0.41$\pm$0.26 & 0.07$\pm$0.06 &  ...  \\
                         & {\bf  0.64}   & {\bf  0.62}   & {\bf  0.98}
& {\bf  0.40} &{\bf  0.07} &  {\bf ... }\\
$[$OIII$]$ $\lambda$5007 & 1.92$\pm$0.08 & 1.89$\pm$0.08 &
2.09$\pm$0.71 & 0.99$\pm$0.58 &0.25$\pm$0.12 & 0.28$\pm$0.12\\
                         & {\bf  1.85}   & {\bf  1.84}   & {\bf  1.95}
& {\bf  0.81} &{\bf  0.23} & {\bf  0.26} \\ 
HeI $\lambda$5876        & 0.20$\pm$0.03 & 0.17$\pm$0.03 & ...
& ... & 0.16$\pm$0.08 &  0.09$\pm$0.04\\
                         & {\bf  0.16}   & {\bf  0.14}   & {\bf ... }
& {\bf  ...} &{\bf  0.10} & {\bf  0.06} \\ 
$[$OI$]$ $\lambda$6300   & 0.09$\pm$0.03 & 0.09$\pm$0.03 & ...
& 1.34$\pm$0.84 &0.18$\pm$0.06 & 0.09$\pm$0.05\\ 
                         & {\bf  0.06}   & {\bf  0.07}   & {\bf ... }
& {\bf  0.81} &{\bf  0.10} & {\bf  0.05} \\
$[$NII$]$ $\lambda$6548  & 0.31$\pm$0.05 & 0.33$\pm$0.06 &
0.27$\pm$0.23  & 1.11$\pm$0.85 & 0.49$\pm$0.20 &  0.39$\pm$0.18\\
                         & {\bf  0.21}   & {\bf  0.25}   & {\bf  0.13}
& {\bf  0.63} &{\bf  0.24} & {\bf  0.19} \\ 
H$\alpha$                & 4.15$\pm$0.17 & 3.80$\pm$0.15 &
5.77$\pm$1.62 & 5.05$\pm$2.17 &5.76$\pm$0.86 &5.94$\pm$1.07\\
                         & {\bf  2.85}   & {\bf  2.85}   & {\bf  2.85}
& {\bf  2.85} & {\bf  2.85} & {\bf  2.85} \\
$[$NII$]$ $\lambda$6583  & 0.70$\pm$0.06 & 0.67$\pm$0.06 &
0.80$\pm$0.34 & 2.56$\pm$1.33 & 1.86$\pm$0.35 & 1.20$\pm$0.34\\
                         & {\bf  0.48}   & {\bf  0.50}   & {\bf  0.39}
& {\bf  1.44} &{\bf  0.91} & {\bf  0.57} \\
$[$SII$]$ $\lambda$6717  & 0.53$\pm$0.10 & 0.44$\pm$0.07 &
0.95$\pm$0.45  & 1.36$\pm$0.95 & 1.00$\pm$0.28 & 0.69$\pm$0.24\\
                         & {\bf  0.35}   & {\bf  0.32}   & {\bf  0.45}
& {\bf  0.74} &{\bf  0.47} & {\bf  0.32} \\
$[$SII$]$ $\lambda$6731  & 0.42$\pm$0.10 & 0.35$\pm$0.06 &
0.98$\pm$0.44  & 1.06$\pm$0.86 &0.73$\pm$0.22 & 0.60$\pm$0.23\\
                         & {\bf  0.28}   & {\bf  0.26}   & {\bf  0.46}
& {\bf  0.57} & {\bf  0.34} & {\bf  0.27} \\
 & & & & & & \\
F(H$\beta$)              & 89.0$\pm$3.0  & 67.16$\pm$2.01& 9.4$\pm$2.1
& 14.2$\pm$5.0 & 16.30$\pm$2.12 & 28.0$\pm$4.0 \\
                         & {\bf 277.0}    & {\bf  160.6}  & {\bf 79.9}
& {\bf 80.0} &{\bf 137.4} & {\bf 261.0}\\
E(B-V)                   & 0.34$\pm$0.04 & 0.17$\pm$0.03 &
0.65$\pm$0.26 & 0.52$\pm$0.39 & 0.56$\pm$0.14 & 0.67$\pm$0.16 \\
\enddata
\tablecomments{Bold-face numbers are extinction-corrected intensities. 
All fluxes and intensities are relative to
H$\beta$; F(H$\beta$) is in units of 10$^{-16}$ erg cm$^{-2}$ s$^{-1}$.
Spectra 1$\ast$, 1, and 1a are all referred to galaxy 1 (see section 4 for further
explanations); spectra 4 and 4$\ast$ are centered on galaxy 4 at P.A. = 130\degr\ and
P.A. = 90\degr, respectively.}
\end{deluxetable}


\begin{deluxetable}{cccccccc}
\tabletypesize{\footnotesize}
\tablecaption{Fluxes and Intensities of TDG Candidates in CG~J1720-67.8 \label{tdgflux}}
\tablewidth{0pt}
\tablehead{
\colhead{Line} & \colhead{Obj. 3} & \colhead{Obj. 7} & \colhead{Obj. 8} & \colhead{Obj. 9} &
\colhead{Obj. 10}}
\startdata
$[$OII$]$ $\lambda$3727  & ... &  4.19$\pm$0.80 & 3.80$\pm$1.52 &
3.93$\pm$1.65 & 2.55$\pm$1.45 \\
                         &{\bf ...} & {\bf  6.06} & {\bf  4.37} & {\bf
6.32} & {\bf  2.57} \\
H$\gamma$                & ... &0.48$\pm$0.18 & ... & ... & ...\\
                         &{\bf ...} &{\bf  0.58} & {\bf ...} &{\bf ...} &{\bf ...}\\
$[$OIII$]$ $\lambda$4959 &  0.56$\pm$0.50 &  0.85$\pm$0.24 & 
1.02$\pm$0.62 & ... & ...\\
                         &{\bf  0.53} & {\bf  0.83} & {\bf  1.01} & {\bf ...} &{\bf ...}\\
$[$OIII$]$ $\lambda$5007 &1.09$\pm$0.62 &2.43$\pm$0.49 & 1.86$\pm$0.76 
& 1.81$\pm$1.00 & 0.59$\pm$0.41 \\
                         &{\bf  1.00} & {\bf  2.33} & {\bf  1.83} &
{\bf  1.71} & {\bf  0.59} \\ 
$[$NII$]$ $\lambda$6548  &0.72$\pm$0.64 & 0.27$\pm$0.29 &
0.28$\pm$0.43 & 0.38$\pm$0.64 & ...\\
                         &{\bf  0.31} & {\bf  0.17} & {\bf  0.24} &
{\bf  0.22} & {\bf ... }\\
H$\alpha$                & 6.67$\pm$3.07 & 4.43$\pm$0.84 &
3.37$\pm$1.35 & 5.03$\pm$1.81 & 2.88$\pm$1.44\\
                         &{\bf  2.85} & {\bf  2.85} & {\bf  2.85} &{\bf  2.85} &
{\bf  2.85} \\ 
$[$NII$]$ $\lambda$6583  &1.68$\pm$1.01 &0.76$\pm$0.35 & 0.85$\pm$0.59 
& 1.13$\pm$0.85 & 0.39$\pm$0.37 \\
                         &{\bf  0.71} &{\bf  0.49} & {\bf  0.72} &
{\bf  0.64} & {\bf  0.39} \\ 
$[$SII$]$ $\lambda$6717  &2.37$\pm$2.37 & 1.06$\pm$0.39 &
0.74$\pm$0.44 & 1.70$\pm$0.93 & ...  \\ 
                         &{\bf  0.96} &{\bf  0.66} & {\bf  0.62} &
{\bf  0.93} & {\bf ... }\\ 
$[$SII$]$ $\lambda$6731  &1.42$\pm$1.85 & 0.92$\pm$0.38 &0.64$\pm$0.44 
& 1.02$\pm$0.65 &  ...  \\ 
                         &{\bf  0.57} & {\bf  0.57} & {\bf  0.53} &
{\bf  0.55} & {\bf ... }\\ 
 & & & & & \\
F(H$\beta$)              & 7.9$\pm$3.2 & 13.0$\pm$2.0 &7.0$\pm$2.0 &
3.5$\pm$0.9 & 8.9$\pm$3.0\\
                         &{\bf 104.0} & {\bf 48.0 } & {\bf 11.7} &
{\bf 19.3} & {\bf 9.2}\\ 
E(B-V)                   & 0.78$\pm$0.42 & 0.40$\pm$0.17 &
0.15$\pm$0.36 & 0.52$\pm$0.33 & 0.01$\pm$0.67 \\
\enddata
\tablecomments{Bold-face numbers are extinction-corrected intensities. 
All the fluxes and intensities are relative to
H$\beta$; F(H$\beta$) is in units of 10$^{-16}$ erg cm$^{-2}$ s$^{-1}$.}
\end{deluxetable}


\begin{deluxetable}{lccccccccccc}
\tabletypesize{\footnotesize}
\tablecaption{Apparent Magnitudes and Colors\label{mag}}
\tablewidth{0pt}
\tablehead{
\colhead{Band} & \multicolumn{3}{c}{Galaxy 1}& &\multicolumn{3}{c}{Galaxy 2} & &
\multicolumn{3}{c}{Galaxy 4} \\
 &\colhead{(1)} &\colhead{(2)} &\colhead{(3)}& &\colhead{(1)} &
\colhead{(2)} &\colhead{(3)}& &\colhead{(1)} &\colhead{(2)} & \colhead{(3)}}
\tablecolumns{12}
\startdata
B & 17.69& 0.34& 17.36&& 17.35& 0.89& 16.46&& 16.45& 0.47& 15.97\\
V & 17.17& 0.29& 16.89&& 16.37& 0.80& 15.58&& 15.81& 0.39& 15.42\\
R & 16.84& 0.26& 16.58&& 15.87& 0.75& 15.12&& 15.26& 0.35& 14.91\\
J & 15.06& 0.08& 14.98&& 14.04& 0.24& 13.81&& 13.10& 0.12& 13.88\\
H & 14.60& 0.05& 14.54&& 13.36& 0.15& 13.21&& 13.48& 0.07& 13.40\\
K$\rm _s$ & 14.49& 0.03& 14.46&&13.15& 0.10& 13.06&& 13.27& 0.05& 13.22\\
 & & & && & && & & & \\
B$-$V &  0.52&  0.05& 0.47&& 0.98& 0.10& 0.88&& 0.64& 0.08& 0.55\\
V$-$R &  0.33&  0.03& 0.30&& 0.50& 0.05& 0.46&& 0.55& 0.04& 0.51\\
J$-$H &  0.47&  0.03& 0.43&& 0.68& 0.09& 0.59&& 0.52& 0.04& 0.48\\
H$-$K$\rm _s$ & 0.11& 0.02& 0.09&& 0.21& 0.05& 0.16&& 0.21& 0.03& 0.18\\
J$-$K$\rm _s$ & 0.57& 0.05& 0.52&& 0.89& 0.14& 0.75&& 0.73& 0.07& 0.66\\
V$-$K$\rm _s$ & 2.68& 0.25& 2.43&& 3.22& 0.70& 2.52&& 2.54& 0.34& 2.20\\
\enddata
\tablecomments{For each galaxy, col. (1) lists observed magnitudes and colors after Galactic extinction
correction (extinction values from NED, based on Schlegel et al. (1998)); col. (2) lists internal
extinctions and corresponding color excesses derived as a function of morphological type and disk 
inclination according to RC3; col. (3) lists magnitudes corrected for Galactic and internal extinction.}
\end{deluxetable} 


\begin{deluxetable}{lcccccccccccccccccccc}
\tabletypesize{\scriptsize}
\tablecaption{GIM2D Best-Fit Parameters\label{gim2dparam}}
\rotate
\tablewidth{0pt}
\tablehead{
 & \multicolumn{6}{c}{Galaxy 1} & &\multicolumn{6}{c}
{Galaxy 2} & & \multicolumn{6}{c}{Galaxy 4} \\
\colhead{Param.} & \colhead{B} & \colhead{V} & \colhead{R} & \colhead{J} & \colhead{H} &
\colhead{K$_{\rm s}$} && 
\colhead{B} & \colhead{V} & \colhead{R}& \colhead{J} & \colhead{H} &
\colhead{K$_{\rm s}$}&& \colhead{B} & \colhead{V} & \colhead{R}& \colhead{J} & \colhead{H} &
\colhead{K$_{\rm s}$}}
\startdata
B/T               &1.00\tablenotemark{a} &0.44 & 0.56 & 0.67 & 0.71 & 0.71 && 0.71& 0.29& 0.44 & 0.56 & 0.63 & 0.63 && 0.19& 0.22& 0.29 & 0.47 & 0.57 & 0.58 \\
$r_e$ (arcsec)    & 2.69& 1.23& 2.18 & 0.21 & 0.19 & 0.04 && 1.89& 0.21& 0.45 & 0.55 & 0.59 & 0.53 && 3.50& 4.35& 4.09 & 2.60 & 2.49 & 1.87 \\
$h_D$ (arcsec)    & 1.92& 7.78& 4.83 & 4.52 & 4.15 & 1.61 && 2.15& 1.57& 1.52 & 1.87 & 2.26 & 2.29 && 4.37& 4.20& 4.16 & 4.09 & 4.00 & 4.29 \\
$r_{1/2}$ (arcsec)& 2.69& 6.27& 4.57 & 0.53 & 0.41 & 9.67 && 2.47& 1.82& 1.48 & 1.42 & 1.36 & 1.26 && 6.75& 6.63& 6.34 & 4.96 & 4.34 & 3.87 \\
$e$               &0.19 & 0.53& 0.47 & 0.70 & 0.55 & 0.27 && 0.38& 0.70& 0.56 & 0.26 & 0.32 & 0.39 && 0.70& 0.70& 0.64 & 0.55 & 0.61 & 0.65 \\
$\phi_b$ (\degr)  &36.0 & 72.2& 79.0 & 93.1 & 86.7 & 79.4 && $-$169.4& 169.7& 168.1 & 128.1 & 167.8 & 165.0 && 123.0& 124.4& 128.4 & 121.8 &
124.6 & 122.5 \\
$\phi_d$ (\degr)  &41.8 & 158.7& $-$20.27 & 131.8 & 135.4 & 82.4 && 97.2& 211.2& 212.9 & 71.2 & 88.8 & 102.6 && 175.5& 186.9& 183.6 & 183.7
& 189.1 & 184.3 \\
$i$  (\degr)      &39.6 & 75.5& 68.7 & 68.0 & 68.2 & 64.5 && 60.9& 8.6& 3.6 & 16.7 & 39.3 & 41.1 && 46.0& 43.0& 20.0 & 43.1 & 57.4 & 62.7 \\
R$_T$             & 0.04 & 0.07 & 0.09 & 0.02 & 0.03 & 0.07 && 0.04 & 0.02 & 0.01 & 0.03 & 0.00 &
0.01 && 0.10 & 0.07 & 0.05 & 0.02 & 0.01 & 0.01 \\
R$_A$             & 0.13 & 0.11 & 0.19 & 0.06 & 0.04 & 0.00 && 0.14 & 0.06 & 0.02 & 0.07 & 0.02 &
0.01 && 0.10 & 0.17 & 0.08 & 0.06 & 0.00 & 0.02 \\
$\chi^2$          &2.40 & 1.55& 2.35 & 1.32 & 1.13 & 1.18 && 2.06& 1.53& 1.52 & 1.78 & 1.21 & 1.04 && 3.10& 3.24& 1.77 & 1.25 & 1.13 & 0.99 \\
\enddata
\tablenotetext{a}{A comparably good fit with a S\'ersic bulge gives B/T = 0.4.}
\end{deluxetable}


\begin{deluxetable}{ccclc}
\tabletypesize{\footnotesize}
\tablecaption{Positions and Heliocentric Radial Velocities\label{radvel}}
\tablewidth{0pt}
\tablehead{
\colhead{Object Id.} & \colhead{$\alpha$ (J2000)} & \colhead{$\delta$ (J2000)} & 
\colhead{V$_{\rm hel}$
} & \colhead{Error\tablenotemark{a}} \\
           & \colhead{($^h$ $^m$ $^s$)} & \colhead{($^{\circ}$ $^{\prime}$ $^{\prime\prime}$)}
	   &\colhead{(km s$^{-1}$)} & \colhead{(km s$^{-1}$)}}
\startdata
1 & 17 20 27.8& -67 46 20.8& 13430 $\pm$ 1& 14\\
1a&...        &...         & 13708 $\pm$ 12& 21\\
1b&...        &...         & 13756 $\pm$ 15& 21\\
2 & 17 20 28.8& $-$67 46 25.2& 13558 $\pm$ 4& 14\\
3 & 17 20 29.8& $-$67 46 30.6& 13523 $\pm$ 4& 14\\
4 & 17 20 28.7& $-$67 46 31.4& 13427 $\pm$ 10& 17\\
5 & 17 20 27.9& $-$67 46 27.9& 13641 $\pm$ 19& 21\\
6\tablenotemark{b}&17 20 29.6&$-$67 46 34.3& 13639 $\pm$ 88 & 17\\
7 & 17 20 26.4& $-$67 46 45.7& 13211 $\pm$ 13& 23\\
8 & 17 20 26.0& $-$67 46 43.3& 13197 $\pm$ 11& 17\\
9 & 17 20 29.7& $-$67 46 33.1& 13584 $\pm$ 24& 37\\
10 &17 20 27.8& $-$67 46 42.9& 13154 $\pm$ 20& 31\\
11\tablenotemark{c}& 17 20 25.8& $-$67 46 30.4& 13861 $\pm$ 69& 39\\
\enddata
\tablenotetext{a}{The calibration error is calculated at the H$\alpha$ 
wavelength.}
\tablenotetext{b}{The radial velocity of obj. 6 rests on 
H$\alpha$ and [SII] lines, all having a very low SNR.}
\tablenotetext{c}{The radial velocity of obj. 11 is based on the 
only detectable emission line, which was assumed to be [\ion{O}{2}] $\lambda$3727.}
\end{deluxetable}


\begin{deluxetable}{cccc}
\tabletypesize{\footnotesize}
\tablecaption{L$_{\rm H\alpha}$ and SFR \label{sfr}}
\tablewidth{0pt}
\tablehead{
\colhead{Spectrum Id.}& \colhead{L$_{\rm H\alpha}$} & \colhead{SFR} & \colhead{SFRD}\\
 & (10$^{40}$ erg s$^{-1}$) & (M$_{\sun}$ yr$^{-1}$) & (10$^{-8}$ M$_{\sun}$ yr$^{-1}$ 
pc$^{-2}$) }
\startdata
1* & 30.82 & 2.18& 21.22\\
1a & 8.89& 0.63& 11.97\\
2 & 8.93& 0.63& 11.03\\
4 & 15.26& 1.08& 9.98\\
4*& 28.97& 2.05& 19.51\\
\enddata
\end{deluxetable}


\begin{deluxetable}{ccccc}
\tabletypesize{\footnotesize}
\tablecaption{Photoionization Model Parameters\label{modelparam}}
\tablewidth{0pt}
\tablehead{
\colhead{Model Param.}  & \colhead{Obj. 1}& \colhead{Obj. 1a}& \colhead{Obj. 2}& \colhead{Obj. 4}}
\startdata
T$_{\ast}$  (10$^4$ K)&4.0 &4.0& 4.0 &4.0 \\
Z/Z$_{\odot}$ &0.18 &0.18&0.45  &0.08 \\
N/N$_{\odot}$ &0.14 &0.14&0.45  &0.08 \\
O/O$_{\odot}$ &0.16 &0.16&0.58  &0.08 \\
S/S$_{\odot}$ &0.18 &0.18&0.45  &0.09 \\
U (10$^{-4}$)&6.98 & 7.50&2.49  &2.21 \\
N$_{\rm H}$ (10$^2$cm$^{-3}$) &1.35 &1.33& 2.47 &3.12 \\
\enddata
\end{deluxetable}


\begin{deluxetable}{ccccc}
\tabletypesize{\footnotesize}
\tablecaption{Observed and Modeled Line Intensities for the Three Main Galaxies\label{photmodel}}
\tablewidth{0pt}
\tablehead{
\colhead{Line } & \colhead{Obj. 1} &\colhead{ Obj. 1a}& \colhead{Obj. 2}&\colhead{Obj. 4}}
\startdata
$[$OII$]$ 3727 &3.26$\pm$0.15& 3.70$\pm$0.34 & 6.23$\pm$0.59 & 1.70$\pm$0.33 \\
 &{\bf 3.62, 5.6E-1 } &{\bf 3.54, 1.8E-2} &{\bf 5.07, 1.5E-1} &{\bf 2.58, 2.5E+0}\\
H$\delta$ &0.48$\pm$0.46&... &... &... \\
 &{\bf 0.27, 3.0E+0 } &{\bf... } &{\bf... } &{\bf... } \\
H$\gamma$ &0.49$\pm$0.14 &0.53$\pm$0.61 &... &...  \\
 &{\bf 0.48, 4.0E-2 } &{\bf 0.48, 3.4E-2} &{\bf... } &{\bf... } \\
$[$OIII$]$ 4959 &0.64$\pm$0.08 &0.98$\pm$0.46 &0.40$\pm$0.63 &0.08$\pm$0.46  \\
 &{\bf 0.64, 2.4E-3 } &{\bf 0.69, 8.1E-1 } &{\bf 0.18, 3.6E+0}&{\bf 0.08, 1.3E-3 } \\
$[$OIII$]$ 5007 &1.85$\pm$0.04 &1.95$\pm$0.34 &0.94$\pm$0.59 &0.26$\pm$0.43  \\
 &{\bf 1.85, 4.3E-3 } &{\bf 2.21, 6.1E-3 } &{\bf 0.53, 1.8E+0 }&{\bf 0.23, 1.1E-1 } \\
HeI 5876 &0.16$\pm$0.15 &... &... &0.06$\pm$0.44 \\
 &{\bf 0.13, 1.7E+0 } &{\bf ... } &{\bf... } &{\bf 0.13, 8.1E0} \\
$[$OI$]$ 6300 &0.06$\pm$0.33 &... &0.81$\pm$0.63 &0.05$\pm$0.55 \\
 &{\bf 0.05, 6.4E-1 } &{\bf ... } &{\bf 0.18, 3.2E+1 } &{\bf 0.06, 5.4E-2 } \\
$[$NII$]$ 6548 &0.21$\pm$0.13 &0.13$\pm$0.85 &0.63$\pm$0.77 &0.19$\pm$0.46 \\
 &{\bf 0.21, 4.7E-4 } &{\bf 0.20, 4.2E-1 } &{\bf 0.50, 1.2E-1 }&{\bf 0.19, 6.9E-4 } \\
H$\alpha$ &2.85$\pm$0.04 &2.85$\pm$0.28 &2.85$\pm$0.43 &2.85$\pm$0.18 \\
 &{\bf 2.85,5.0E-5 } &{\bf 2.85, 1.2E-10 } &{\bf 2.87, 3.1E-4 }&{\bf 2.89, 6.7E-3 }  \\
$[$NII$]$ 6583 &0.48$\pm$0.08 &0.39$\pm$0.43 &1.44$\pm$0.52&0.57$\pm$0.28 \\
 &{\bf 0.61, 1.3E+1} &{\bf 0.60, 1.5E+0 } &{\bf 1.46, 9.1E-4} &{\bf 0.57, 2.6E-4}  \\
$[$SII$]$ 6716 &0.35$\pm$0.18 &0.45$\pm$0.47 &0.74$\pm$0.70 &0.32$\pm$0.35 \\
 &{\bf 0.34, 2.2E-2 } &{\bf 0.33, 6.2E-1 } &{\bf 0.75, 1.1E-4 }&{\bf 0.30, 3.9E-2 }  \\
$[$SII$]$ 6731 &0.28$\pm$0.24 &0.46$\pm$0.45 &0.57$\pm$0.81 &0.27$\pm$0.38 \\
 &{\bf 0.27, 5.3E-2 } &{\bf 0.26, 3.2E+0 } &{\bf 0.62, 1.4E-2 }&{\bf 0.26, 1.6E-2 } \\
\enddata
\tablecomments{Line intensities are relative to H$\beta$. For every emission
line, the observed values with \emph{relative} errors are in the first row and 
the modeled values with the $\chi^2$ of the
fit to the line are in the second row, in bold type.}
\end{deluxetable}

\end{document}